\documentclass[aps,onecolumn,superscriptaddress,nofootinbib,amsmath,amssymb,floatfix]{revtex4}
\usepackage{graphicx}  
\usepackage{dcolumn}   
\usepackage{bm,color}        
\usepackage{amssymb}   
\usepackage{adjustbox}
\usepackage{hyperref}

\definecolor{MyBlue}{rgb}{0.15,0.15,0.70}

\hypersetup{
colorlinks=true,
citecolor=red,
linkcolor=MyBlue,
urlcolor=MyBlue
}

\newcommand{\be}{\begin{equation}}
\newcommand{\ee}{\end{equation}}
\newcommand{\beq}{\begin{equation}}
\newcommand{\eeq}{\end{equation}}
\newcommand{\bea}{\begin{eqnarray}}
\newcommand{\eea}{\end{eqnarray}}

\newcommand{\iu}{{i\mkern1mu}}

\def\pa{\partial}

\def\k{{\bf k}}

\def\x{{\bf x}}

\newcommand\ees{\end{eqnarray}}
\newcommand\bees{\begin{eqnarray}}

\def\pa{\partial}

\graphicspath{Plots/}

\begin{document}

\pagenumbering{arabic}

\title{In search of an observational quantum signature of the primordial perturbations \\ in slow-roll and ultra slow-roll inflation} 
\author{Roland de Putter}
\address{California Institute of Technology, Pasadena, CA}
\author{Olivier Dor{\'e}}
\email{olivier.p.dore@jpl.nasa.gov}
\address{Jet Propulsion Laboratory, California Institute of Technology, Pasadena, CA}
\address{California Institute of Technology, Pasadena, CA}

\begin{abstract}
In the standard inflationary paradigm, cosmological density perturbations are generated as quantum fluctuations in the early Universe, but then undergo a quantum-to-classical transition.
A key role in this transition is played by squeezing of the quantum state,
which is a result of the strong suppression of the decaying mode component of the perturbations.
Motivated by ever improving measurements of the cosmological perturbations,
we ask
whether there are scenarios where this decaying mode is nevertheless still observable in the late Universe, ideally leading to a ``smoking gun'' signature of the quantum nature of the perturbations.
We address this question by evolving the quantum state of the perturbations from inflation into the post-inflationary Universe.
After recovering the standard result that in slow-roll (SR) inflation the decaying mode is indeed hopelessly suppressed by the time the perturbations are observed (by $\sim 115$ orders of magnitude), we turn to ultra slow-roll (USR) inflation, a scenario in which the usual decaying mode actually grows on super-horizon scales.
Despite this drastic difference in the behavior of the mode functions, we find also in USR that the late-Universe decaying mode amplitude is dramatically suppressed, in fact by the same $\sim 115$ orders of magnitude.
We finally explain that this large suppression is a general result that holds beyond the SR and USR scenarios considered and follows from a modified version of Heisenberg's uncertainty principle and the observed amplitude of the primordial power spectrum.
The classical behavior of the perturbations is thus closely related to the classical behavior of macroscopic objects drawing an analogy with the position of a massive particle, the curvature perturbations today
have an enormous effective mass of order $m_{\rm pl}^2/H_0^2 \sim 10^{120}$, making them highly classical.




\end{abstract}

\maketitle

\section{Introduction}


Inflation describes a phase of accelerated expansion in the very first moments of the Universe~\cite{mukhbook, baumanntasi09}. As a theoretical paradigm, inflation has been increasingly supported by ever more discriminating data: a small spatial curvature, and Gaussian, adiabatic, primordial perturbations with a nearly scale-invariant power spectrum have been measured \cite{Akrami2018-pr}. Nevertheless, the correctness of this paradigm remains debated and, arguably, an unambiguous observational proof remains elusive \cite{Chowdhury2019-jz}. This work aims to contribute to the quest for such an observational proof.

Initially invented to circumvent difficulties in the Friedmann-Lema\^itre-Robertson-Walker model, inflation also provides a mechanism to generate the primordial density perturbations that grew to form the large-scale structure we observe today. Importantly, in the inflation paradigm, these perturbations are vacuum quantum fluctuations amplified by gravitational instability and stretched over cosmological distances. The perturbations we observe today are of quantum origin. To observationally establish this quantum origin in a direct manner would prove inflation and surely suggest new observational connections to quantum gravity. 

These prospects have motivated studies of the quantum nature of inflationary perturbations since inflation was proposed~\cite{Starobinsky:1986fx,Grishchuk:1990bj,albrechtetal94,Polarski:1995jg,Lesgourgues:1996jc,Egusquiza:1997ez, Kiefer:1998qe, Perez:2005gh, Campo:2005sv, Ellis:2006fy,Kiefer:2006je, Kiefer:2008ku,Valentini:2008dq,Koksma:2010dt, Bassi:2010ss, PintoNeto:2011ui, Martin:2012pea, Canate:2012ua, Lochan:2012di, Bassi:2012bg, Das:2013qwa, Oriti:2013jga, Markkanen:2014dba, Das:2014ada, Maldacena:2015bha, Banerjee:2015zua, Singh:2015sua, Vitenti:2015lpa, Goldstein:2015mha, Leon:2015hwa, Leon:2015ija, Colin:2015tla, Valentini:2015sna,Lim2015-uu,Martin:2015qta,Martin:2016nrr,Martin:2016tbd,Martin:2017zxs,Martin:2018lin,Martin:2018zbe, nelson16, boddyetal16}. Recently, novel concepts borrowed from quantum information theory such as the \emph{quantum discord} were invoked to capture the quantumness of inflationary perturbations~\cite{Lim2015-uu,Martin:2015qta,Martin:2018lin}, and the connection to Bell inequalities on cosmological scales was articulated and studied~\cite{Campo:2005sv,Maldacena:2015bha,Martin:2016tbd,choudhury16,Martin:2017zxs,Martin2019-op}. 

In this work, we revisit how the quantumness of these perturbations appears, how it evolves in the quantum-to-classical transition and whether there exist quantum relics we could measure. We will focus on the squeezing\footnote{The other process key to any understanding of the quantum-to-classical transition is decoherence due to interactions with the ``environment'', which we briefly discuss at the end of Section \ref{sec:slow-roll} and of Section \ref{sec:Heis}.} of the quantum state that occurs during the quantum-to-classical transition~\cite{Grishchuk:1990bj}. Since this squeezing is associated with a strong suppression of the decaying mode, a measurement of this decaying mode would be related to the quantumness of the initial perturbations (see \cite{Kodwani:2019ynt} and references therein for a recent discussion). We thus ask whether there exist scenarios where this decaying mode could be less suppressed and potentially measurable. We find that it is not the case and that the suppression is a general result that holds beyond the specific scenarios considered. We explain how it follows from a modified version of Heisenberg's uncertainty principle and the observed amplitude of the primordial power spectrum.

We will review in Sec.~\ref{sec:slow-roll} the primordial quantum fluctuations in slow-roll inflation,  detailing in particular the squeezing stage and the resulting classical behavior of the associated perturbations. Motivated by the search for a remaining quantum signature, we discuss in Sec.~\ref{sec:SR2RD} the evolution of the quantum state after slow-roll inflation (i.e.~into the late Universe) before discussing quantum signatures in ultra-slow roll inflation in Sec.~\ref{sec:USR}. Inspired by this example, we discuss in Sec.~\ref{sec:Heis} how a generalized version of Heisenberg's uncertainty principle explains the general suppression of quantum signatures.


\section{Primordial quantum fluctuations in slow-roll inflation}
\label{sec:slow-roll}

\subsection{The action}

Unless otherwise stated, we work in Planck units, setting in particular $c = \hbar = 1$.
For a perfect fluid, the action in terms of the comoving curvature perturbation, $\mathcal{R}$ \cite{bardeen80,BST83,lyth85}, is to second order given by \cite{mukhbrandfeld92,garmukh99},
\bea
\label{eq:action}
S &=& \frac{1}{2} \int d\tau \, d^3 x \, \frac{2 \epsilon \, m_{\rm pl}^2}{c_s^2} \, a^2 \, \left( {\mathcal{R}'}^2 - c_s^2 \, \left( \pa_i \mathcal{R} \right)^2\right) \nonumber \\
&=&  \int d\tau \,  \sum_\k \frac{1}{2} \, z^2(\tau) \, \left( {\mathcal{R}'}_\k \, {\mathcal{R}'}_{- \k} - c_s^2 \, k^2 \, \mathcal{R}_\k \, \mathcal{R}_{- \k} \right),
\eea
where $a$ is the scale factor,
\beq
\epsilon \equiv -\dot{H}/H^2 = \frac{3 (1 + w)}{2},
\eeq
with $w = p/\rho$ the equation of state and a dot indicating a derivative with respect to coordinate time $t$, $c_s$ is the sound speed, $m_{\rm pl} \equiv (8 \pi G)^{-1/2}$ the reduced Planck mass, and we have defined the combination,
\beq
z(\tau) \equiv \sqrt{\frac{2 \epsilon \, m_{\rm pl}^2}{c_s^2}} \, a(\tau).
\eeq
The spatial coordinates in Eq.~(\ref{eq:action}) are comoving coordinates, $\tau = dt/a$ is conformal time, and
primes denote derivatives with respect to $\tau$.
The action (\ref{eq:action}) in particular describes the curvature perturbations during single-field inflation\cite{mukh85,sasaki86,mukhbrandfeld92,Maldacena:2002vr}. In the second line of Eq.~(\ref{eq:action}), we have expressed $\mathcal{R}$ in Fourier space, using the Fourier convention\footnote{We here choose the slightly unconventional finite-volume Fourier convention in order to avoid delta functions in the commutation relations that follow and thus to make the analogy with the simple harmonic oscillator even more direct.},
\beq
\mathcal{R}(\x) = \frac{1}{V^{1/2}} \, \sum_{\k} e^{- \iu \k \cdot \x} \, \mathcal{R}_\k.
\eeq
The action above thus describes a set of independent\footnote{More precisely, each pair $\mathcal{R}_\k$, $\mathcal{R}_{-\k}$ describes an independent set of two real-valued variables.} Fourier modes $\mathcal{R}_\k$, with equation of motion,
\beq
\label{eq:EoM}
\left( z^2(\tau) \, {\mathcal{R}'}_\k \right)' + c_s^2 \, k^2 \, z^2(\tau) \, \mathcal{R}_\k = 0.
\eeq

\subsection{Quantization}

The conjugate momenta to the curvature perturbations are given by,
\beq
\Pi_\k = \frac{\pa S}{\pa \mathcal{R}'_{-\k}} = z^2(\tau) \, \mathcal{R}'_{\k}.
\label{eq:pi def}
\eeq
The perturbations are now quantized by promoting $\mathcal{R}_\k$ and $\Pi_\k$ to operators (marked throughout this paper by $\hat{}$) and by imposing the canonical commutation relations,
\beq
\left[ \hat{\mathcal{R}}_\k, \hat{\Pi}_{-\k} \right] = \iu, \quad \text{etc.}
\eeq
These operators are time-dependent in the Heisenberg picture.
We make this time dependence explicit by
expanding $\hat{\mathcal{R}}_\k$ in terms of positive and negative frequency solutions of the equations of motion, $f_k(\tau)$ and $f_k^*(\tau)$,
\beq
\label{eq:def creat anni}
\hat{\mathcal{R}}_\k(\tau) = f_k(\tau) \, \hat{a}_\k + f_k^*(\tau) \, \hat{a}^\dagger_{-\k}.
\eeq
Here the operators $\hat{a}^\dagger_{-\k}$ and $\hat{a}_\k$ are time-independent
and we enforce that they satisfy
the canonical commutation relations,
\beq
\left[ \hat{a}_\k, \hat{a}_\k^\dagger \right] = 1, \quad \text{etc.},
\eeq
by demanding that the conserved Wronskian of the solutions $f_k(\tau)$ and $f_k^*(\tau)$ satisfies the normalization,
\beq
\label{eq:Wronskian}
W(f_k,f_k^*) \equiv -\frac{\iu}{2}  \, \left( f_k(\tau) \, z^2(\tau) f_k^{* \, '}(\tau) - f_k^*(\tau) \, z^2(\tau) f'_k(\tau)\right) = \frac{1}{2}.
\eeq
The choice of mode function $f_k(\tau)$ then fixes the Fock space defined by the creation and annihilation operators $\hat{a}_\k$ and $\hat{a}_\k^\dagger$.

\subsection{Slow-roll inflation}
\label{subsec:sri}

While our discussion so far applied to any action of the perfect fluid form, Eq.~(\ref{eq:action}), we now focus on the case of slow-roll, single-field inflation. To describe the slow-roll phase, we assume a background arbitrarily close to de Sitter, with constant Hubble parameter $H_I$ and scale factor,
\beq
a(\tau) = \frac{-1}{H_I \, \tau},
\eeq
and small and constant slow-roll parameter, $\epsilon \ll 1$. We will in practice also assume a canonical kinetic term for the inflaton so that $c_s = 1$, but we will show expressions for general $c_s$.
In future sections we will use the action (\ref{eq:action}) to describe different scenarios, such as ultra slow-roll inflation and a radiation dominated Universe.

In the slow-roll scenario, the equation of motion (\ref{eq:EoM}) has the two independent, real-valued solutions,
\bea
\label{eq:sols SR norm}
{\mathcal{R}}_{{\rm grow},k}^{\rm SR}(\tau) &\equiv& - \sqrt{\frac{\pi}{2}} \, x^{3/2} \, Y_{3/2}(x) =  \cos x + x \, \sin x ,  \nonumber \\
{\mathcal{R}}_{{\rm dec},k}^{\rm SR}(\tau)  &\equiv&   - \sqrt{\frac{\pi}{2}} \, x^{3/2} \, J_{3/2}(x) =  - \sin x + x \, \cos x , \quad \text{with} \quad x \equiv -c_s \, k \, \tau
\eea
where $J_\nu$ and $Y_\nu$ are Bessel function of the first and second kind, respectively.
The solutions are named growing and decaying modes because, after horizon exit, $x \ll 1$, the growing-mode solution approaches unity, while the decaying mode approaches zero (see Section \ref{subsec:squeezing}, Figure \ref{fig:SR}).

We now choose the standard Bunch-Davies solution for the mode functions,
\beq
\label{eq:mode BD}
f_k(\tau) = -\frac{\sqrt{\pi} \, H_I}{2 (c_s \, k)^{3/2}} \, \sqrt{\frac{c_s^2}{2 \epsilon \, m_{\rm pl}^2}} \, x^{3/2} \, H_{3/2}^{(1)}(x) = \frac{1}{\sqrt{2 c_s \, k}} \, z^{-1}(\tau) \, e^{\iu x} \, \left[ 1 + \frac{\iu}{x} \right]
,
\eeq
with $H_{\nu}^{(1)} \equiv J_\nu + \iu \, Y_{\nu}$ the Hankel function of the first kind.
With this choice, the state annihilated by all $\hat{a}_\k$ is the standard Bunch-Davies vacuum, i.e.~the lowest-energy eigenstate of the Hamiltonian for modes deep inside the horizon at early times.
To connect more naturally to the real-valued, independent solutions of the equations of motion in the post-inflationary Universe, we will throughout this paper
describe the state in terms of real-valued components of $f_k$,
\beq
\label{eq:f z1z2}
f_k(\tau) =  \mathcal{R}_{k,2}(\tau) + \iu \, \mathcal{R}_{k,1}(\tau).
\eeq
The solution during slow-roll inflation, Eq.~(\ref{eq:mode BD}), then corresponds to the growing and decaying solutions,
\beq
\label{eq:sols SR}
\mathcal{R}_{k,1}(\tau) = a_\mathcal{R} \, \mathcal{R}^{\rm SR}_{\rm grow}(\tau), \quad \mathcal{R}_{k,2}(\tau) = a_\mathcal{R} \, \mathcal{R}^{\rm SR}_{\rm dec}(\tau),
\eeq
with normalization,
\beq
\label{eq:norm SR}
a_\mathcal{R} \equiv \frac{H_I}{\sqrt{2} \, c_s^{1/2} \, k^{3/2} \, \sqrt{2 \epsilon} \, m_{\rm pl}}.
\eeq
The quantum state of the perturbations after slow-roll inflation is fully characterized by the subsequent evolution of the functions $\mathcal{R}_{k,1}(\tau)$ and $\mathcal{R}_{k,2}(\tau)$.
We have so far given expressions for general sound speed, but from here on we will assume the canonical value, $c_s = 1$.

We note that we are free to multiply $f_k(\tau)$ by an arbitrary phase without changing the spectrum of states defined by $\hat{a}_\k, \hat{a}^\dagger_\k$ and, in particular, without changing the Bunch-Davies state of the primordial perturbations.
This phase change is equivalent to a rotation of the vector of real-valued EoM solutions, $(\mathcal{R}_{k,2}(\tau), \mathcal{R}_{k,1}(\tau))$.
We chose the current basis because, during inflation, it matches the standard growing and decaying mode solutions, Eq.~(\ref{eq:sols SR norm}).
However, we will see in Section \ref{sec:SR2RD} and beyond that, when considering the state of the perturbations after inflation, it may be more convenient to use a rotated basis that corresponds to the standard post-inflationary growing and decaying modes\footnote{As we discuss in Section \ref{sec:SR2RD} and onward, the growing (decaying) mode during inflation does not generally evolve into the exact post-inflationary growing (decaying) mode, but into a linear combination of the two post-inflationary modes.}.

\subsection{Expectation values}
\label{subsec:expvals}

We have now fully defined the initial state and its evolution, which is completely described by the solutions $\mathcal{R}_{k,1}(\tau)$ and $\mathcal{R}_{k,2}(\tau)$ to the classical equation of motion.
The time evolution of the operators $\hat{\mathcal{R}}_\k$ and $\hat{\Pi}_\k$ (defined in Eq.~\ref{eq:pi def}), which follows the classical equations of motion,
can conveniently be written in terms of the (rescaled) initial operators as,
\bea
\label{eq:evol zeta pi}
\hat{\mathcal{R}}_\k(\tau) &=& \sqrt{2} \, \mathcal{R}_{k,2}(\tau) \, \hat{x}_\k - \sqrt{2} \, \mathcal{R}_{k,1}(\tau) \, \hat{p}_\k \nonumber \\
\hat{\Pi}_\k(\tau) &=& \sqrt{2} \, z^2(\tau) \, \mathcal{R}_{k,2}'(\tau) \, \hat{x}_\k - \sqrt{2} \, z^2(\tau) \, \mathcal{R}_{k,1}'(\tau) \, \hat{p}_\k,
\eea
where $\hat{x}_\k = \frac{1}{\sqrt{2}} \, \left( \hat{a}_\k + \hat{a}^\dagger_{-\k} \right)$,
$\hat{p}_\k = -\iu \,  \frac{1}{\sqrt{2}} \, \left( \hat{a}_\k - \hat{a}^\dagger_{-\k} \right)$.
In other words,
the statistics of $\hat{x}_\k$, $\hat{p}_\k$ are
simply those of the position and momentum in the ground state of a simple harmonic oscillator with frequency $\omega = 1$.
In particular, the expectation values of its $2$-point correlators are given by,
\beq
\langle \hat{x}_\k \, \hat{x}_{-\k} \rangle = \langle \hat{p}_\k \, \hat{p}_{-\k} \rangle = \frac{1}{2}, \quad
\langle \hat{x}_\k \, \hat{p}_{-\k} \rangle = - \langle \hat{p}_\k \, \hat{x}_{-\k} \rangle = \frac{\iu}{2}.
\eeq

Eq.~(\ref{eq:evol zeta pi}) is a useful expression for understanding the evolution of expectation values and for gaining an intuitive understanding of the classical limit (see Section \ref{subsubsec:wigner}).
Using this expression,
correlators of $\hat{\mathcal{R}}_\k(\tau)$, $\hat{\Pi}_\k(\tau)$ are easily expressible in terms of the mode functions $\mathcal{R}_{1,2}(\tau)$ (and expectation values of the harmonic oscillator ground state).
The 2-point functions are given by,
\bea
\label{eq:correlations}
\langle \hat{\mathcal{R}}_\k \, \hat{\mathcal{R}}_{-\k} \rangle &=&  \mathcal{R}_{k,1}^2(\tau) + \mathcal{R}_{k,2}^2(\tau) \nonumber \\
\langle \hat{\Pi}_\k \, \hat{\Pi}_{-\k}\rangle &=&  z^4(\tau) \, \left( \mathcal{R}_{k,1}^{' \, 2}(\tau) + \mathcal{R}_{k,2}^{' \, 2}(\tau)\right) \nonumber \\
\langle \hat{\mathcal{R}}_\k \, \hat{\Pi}_{-\k} \rangle &=&  z^2(\tau) \, \left( \mathcal{R}_{k,1}(\tau) \, \mathcal{R}_{k,1}'(\tau) + \mathcal{R}_{k,2}(\tau) \, \mathcal{R}_{k,2}'(\tau) \right) + \frac{\iu}{2} \nonumber \\
\langle \hat{\Pi}_\k \, \hat{\mathcal{R}}_{-\k} \rangle &=& z^2(\tau) \, \left( \mathcal{R}_{k,1}(\tau) \, \mathcal{R}_{k,1}'(\tau) + \mathcal{R}_{k,2}(\tau) \, \mathcal{R}_{k,2}'(\tau) \right) - \frac{\iu}{2}.
\eea

The above quantities directly give the power and cross-spectra of $\mathcal{R}$ and $\Pi$.
\beq
P_\mathcal{R}(k) = \langle \hat{\mathcal{R}}_\k \, \hat{\mathcal{R}}_{-\k} \rangle, \quad
P_\pi(k) = \langle \hat{\Pi}_\k \, \hat{\Pi}_{-\k}\rangle, \quad \text{etc.}
\eeq
With our (standard) power spectrum convention, the dimensionless curvature power spectrum (i.e.~the variance per e-folding in scale $k$) is,
\beq
\Delta^2_\mathcal{R}(k) = \frac{k^3}{2 \pi^2} \, P_\mathcal{R}(k) =
\frac{k^3}{2 \pi^2} \, \left(\mathcal{R}^2_{k,1}(\tau) + \mathcal{R}^2_{k,2}(\tau) \right)
%
\eeq
such that during slow-roll inflation,
\beq
\label{eq:PK modes}
\Delta^2_\mathcal{R}(k) = \frac{k^3 \, a_\mathcal{R}^2}{2 \pi^2} \, \left( \left(\mathcal{R}_{{\rm grow},k}^{{\rm SR}}(\tau)\right)^2
+ \left(\mathcal{R}_{{\rm dec},k}^{{\rm SR}}(\tau)\right)^2 \right)
\to \frac{1}{2 \epsilon \, m_{\rm pl}^2} \, \left(\frac{H_I}{2 \pi}\right)^2 \equiv A_s,
\eeq
where the arrow points to the super-horizon limit where $\mathcal{R}_{{\rm grow},k}^{{\rm SR}} \to 1$ and $\mathcal{R}_{{\rm dec},k}^{{\rm SR}} \to 0$.
The amplitude factor $k^3 \, a_\mathcal{R}^2/(2 \pi^2)$ is thus immediately recognized as the standard expression for the amplitude of the primordial power spectrum in slow-roll inflation, $A_s$ (assuming the expression is evaluated at the appropriate ``pivot scale'', $k_*$).



\subsection{The Wigner function}
\label{subsubsec:wigner}

A convenient description of the quantum state of the primordial perturbations, which we will use throughout this paper, is in terms of the Wigner function (see e.g.~\cite{decoherencebook}).
Before introducing the Wigner function,
note that so far
we have discussed the perturbations in terms of the complex Fourier modes $\mathcal{R}_\k$.
Each pair of complex modes $\mathcal{R}_\k$, $\mathcal{R}_{-\k}$ describes two real-valued modes,
\beq
\label{eq:zetaRI}
\hat{\mathcal{R}}_\k = \frac{1}{\sqrt{2}} \, \left( \hat{\mathcal{R}}_{\k,R} + \iu \, \hat{\mathcal{R}}_{\k,I} \right),
\quad \hat{\Pi}_\k = \frac{1}{\sqrt{2}} \, \left( \hat{\Pi}_{\k,R} + \iu \, \hat{\Pi}_{\k,I} \right).
\eeq
With the above normalization factor of $1/\sqrt{2}$, all expressions for expectation values in terms of complex modes above
can be directly applied to the real degrees of freedom by simply substituting the latter for the former.
For instance,
\beq
\langle \hat{\mathcal{R}}_{\k,R}^2 \rangle = \langle \hat{\mathcal{R}}_{\k,I}^2 \rangle = \langle \hat{\mathcal{R}}_\k \, \hat{\mathcal{R}}_{-\k} \rangle, \quad \text{etc.}
\eeq
For ease of notation, but without loss of generality, we from here on consider real degrees of freedom $\hat{\mathcal{R}}_{\k,R}$, $\hat{\mathcal{R}}_{\k,I}$. We from here on also drop all $k$ subscripts.

In the Schr\"odinger picture, the quantum state of each degree of freedom is described by an evolving Gaussian wave function\footnote{The full wave function is a product of the wave functions of the individual degrees of freedom.} $\psi(\mathcal{R})$, which can be fully specified by the evolution of the mode functions $\mathcal{R}_1(\tau)$ and $\mathcal{R}_2(\tau)$.
The Wigner function (of this pure state) is then defined as,
\beq
W(\mathcal{R}, \Pi) = \frac{1}{\pi \hbar} \, \int dy \, e^{2 \iu \pi \, y} \, \psi^*(\mathcal{R} + y) \, \psi(\mathcal{R} - y),
\eeq
and contains the same information as the wave function.
It has a number of useful properties in common with a phase-space probability distribution, and therefore we refer to it as a pseudo-phase-space distribution.
However, we caution that the Wigner function is {\it not} a true phase-space distribution. First of all, conceptually, a quantum state simply does not have a well-defined phase-space distribution as $\hat{\mathcal{R}}$ and $\hat{\Pi}$ are non-commuting operators.
More concretely, treating $W$ like a phase-space distribution does not in general reproduce the true quantum expectation values, i.e.
\beq
\int d\mathcal{R} \, d\Pi \, W(\mathcal{R}, \Pi) \, A(\mathcal{R}, \Pi) \ne \langle A(\hat{\mathcal{R}}, \hat{\Pi})\rangle,
\eeq
where $A$ is some function of $\mathcal{R}$ and $\Pi$.
In fact, the Wigner function is not even generally positive-definite.

For the primordial fluctuations, the Wigner function is a bivariate Gaussian (which {\it is} positive-definite\footnote{Indeed, Gaussian states are the only pure states with a positive-definite Wigner function.}),
\beq
W(\mathcal{R}, \Pi)
= \frac{1}{\pi} \, \text{Exp}\left\{ -\frac{1}{2} \begin{pmatrix} \mathcal{R} & \Pi \end{pmatrix} \, {\bf C}^{-1} \, \begin{pmatrix} \mathcal{R} \\ \Pi \end{pmatrix} \right\},
\eeq
with covariance matrix,
\beq
\label{eq:covWig}
{\bf C} =\begin{pmatrix} \langle \hat{\mathcal{R}}^2 \rangle & \text{Re}\left(\langle \hat{\mathcal{R}} \, \hat{\Pi} \rangle\right) \\
\text{Re}\left(\langle \hat{\mathcal{R}} \, \hat{\Pi} \rangle\right) & \langle \hat{\Pi}^2 \rangle
\end{pmatrix},
\eeq
where the expectation values are the true quantum expectation values given in Eq.~(\ref{eq:correlations}).
The Wigner function above thus behaves\footnote{This is in general true for the Wigner function of any Gaussian state undergoing linear evolution.} like a Gaussian probability distribution describing the {\it real components} of the true quantum correlations,
\beq
\int d\mathcal{R} \, d\Pi \, W(\mathcal{R}, \Pi) \, A(\mathcal{R}, \Pi) = \text{Re}\left( \langle A(\hat{\mathcal{R}}, \hat{\Pi})\rangle \right), \quad \text{for} \quad A(\hat{\mathcal{R}}, \hat{\Pi}) = \hat{\mathcal{R}}^2, \, \hat{\mathcal{R}} \, \hat{\Pi}, \, \hat{\Pi}^2.
\eeq
Since the operators $\hat{\mathcal{R}}$ and $\hat{\Pi}$ follow the classical equations of motion,
${\bf C}$ evolves as the covariance matrix of a classically evolving stochastic distribution,
and therefore the Wigner function itself evolves as if it were a classical phase-space distribution\footnote{One can show more generally that, for linear equations of motion, the Wigner function obeys the same evolution equations as a classical phase-space distribution.}.
A useful equivalent formulation is that the real components of the cross- and auto-correlations of $\hat{\mathcal{R}}$ and $\hat{\Pi}$
are fully described by the classical evolution in Eq.~(\ref{eq:evol zeta pi}) if we treat $\hat{x}$ and $\hat{p}$ in Eq.~(\ref{eq:evol zeta pi}) as stochastic variables with Gaussian distribution and covariance matrix ${\bf C} = \tfrac{1}{2} \, {\bf 1}_2$. Nevertheless, the Wigner function is still not a true phase-space distribution. First of all, treating it as such does not reproduce the imaginary part of $\langle \mathcal{R} \, \Pi \rangle$.
One might brush off this issue as $\hat{\mathcal{R}} \, \hat{\Pi}$ is not Hermitian and therefore not an observable,
while the expectation value of the Hermitian operator $\tfrac{1}{2} \, \left( \hat{\mathcal{R}} \, \hat{\Pi} +  \hat{\Pi} \, \hat{\mathcal{R}} \right)$ {\it is} reproduced by treating the Wigner function as a phase-space distribution.
However, as soon as we consider higher order operators, the true quantum  expectation values deviate from those obtained from treating the Wigner function as a phase-space distribution even for Hermitian operators (see e.g.~\cite{martinvennin16}).
We will come back to this more quantitatively in Section \ref{subsec:squeezing2classical}.

The Wigner function is however a useful tool for describing the quantum-to-classical transition. One way of defining classical behavior is to require that
the properties of the state of the primordial perturbations can be reproduced by a stochastic phase-space distribution of variables $\mathcal{R}$ and $\Pi$ undergoing classical evolution \cite{Grishchuk:1990bj,Polarski:1995jg}. If this requirement is (approximately) satisfied, based on the above discussion, that phase-space distribution must equal the Wigner function (assuming the above scenario of a Gaussian Wigner function and linear evolution). We can thus quantify the quantum-to-classical transition by comparing the true properties of the quantum state to those computed by treating the Wigner function as a phase-space probability distribution. We will see in the next Section that classicality in the above sense is approached as modes exit the horizon and the Wigner function is {\it squeezed}. We will visualize the Wigner function by its contour of constant $\chi^2 \equiv -2 \ln( \pi W) = 1$ (see e.g.~Figure \ref{fig:SR}), which fully characterizes it because it is a Gaussian.


\subsection{Squeezing upon horizon exit}
\label{subsec:squeezing}

\begin{figure*}[]
\centering
\includegraphics[width=0.47\textwidth]{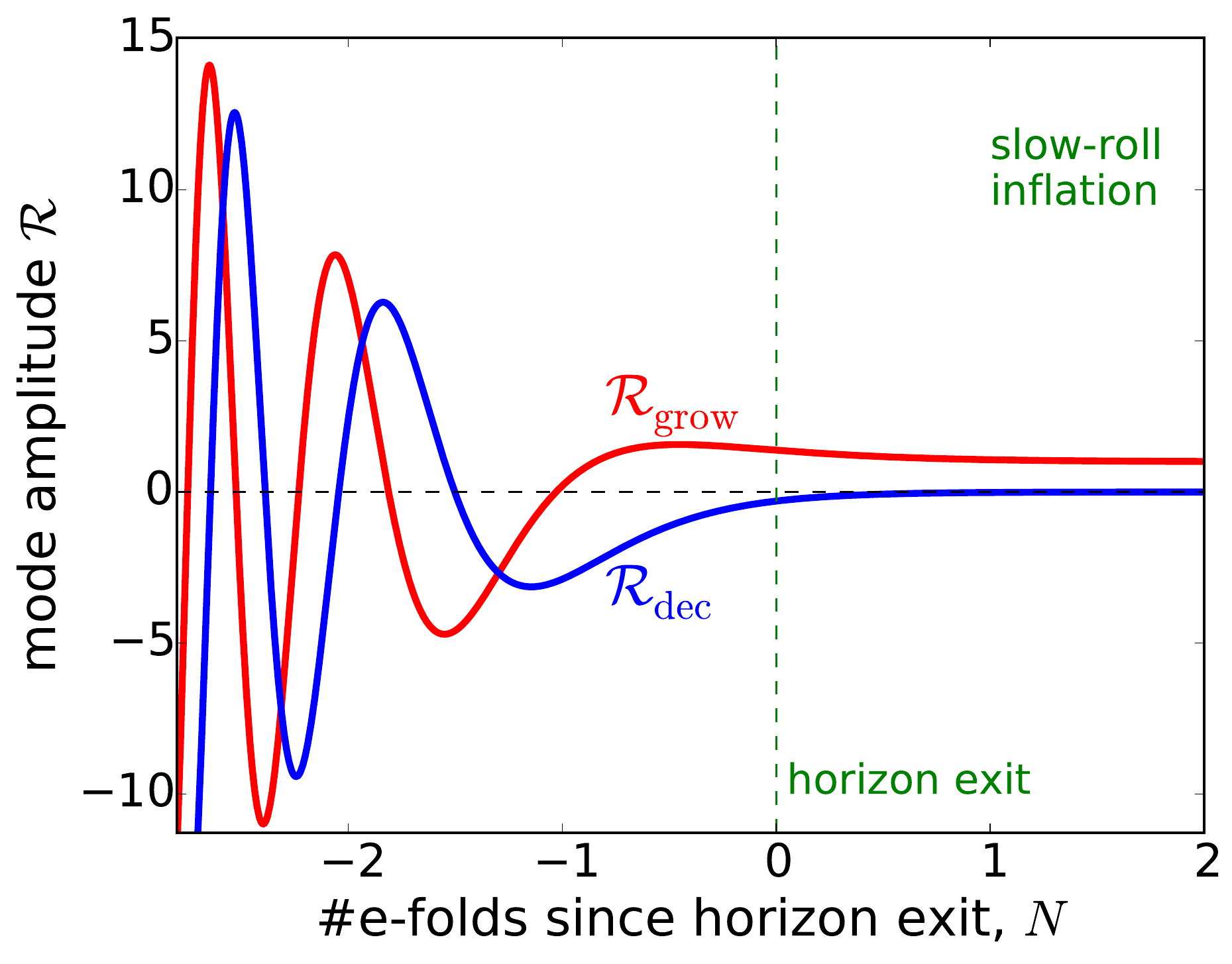} 
\includegraphics[width=0.41\textwidth]{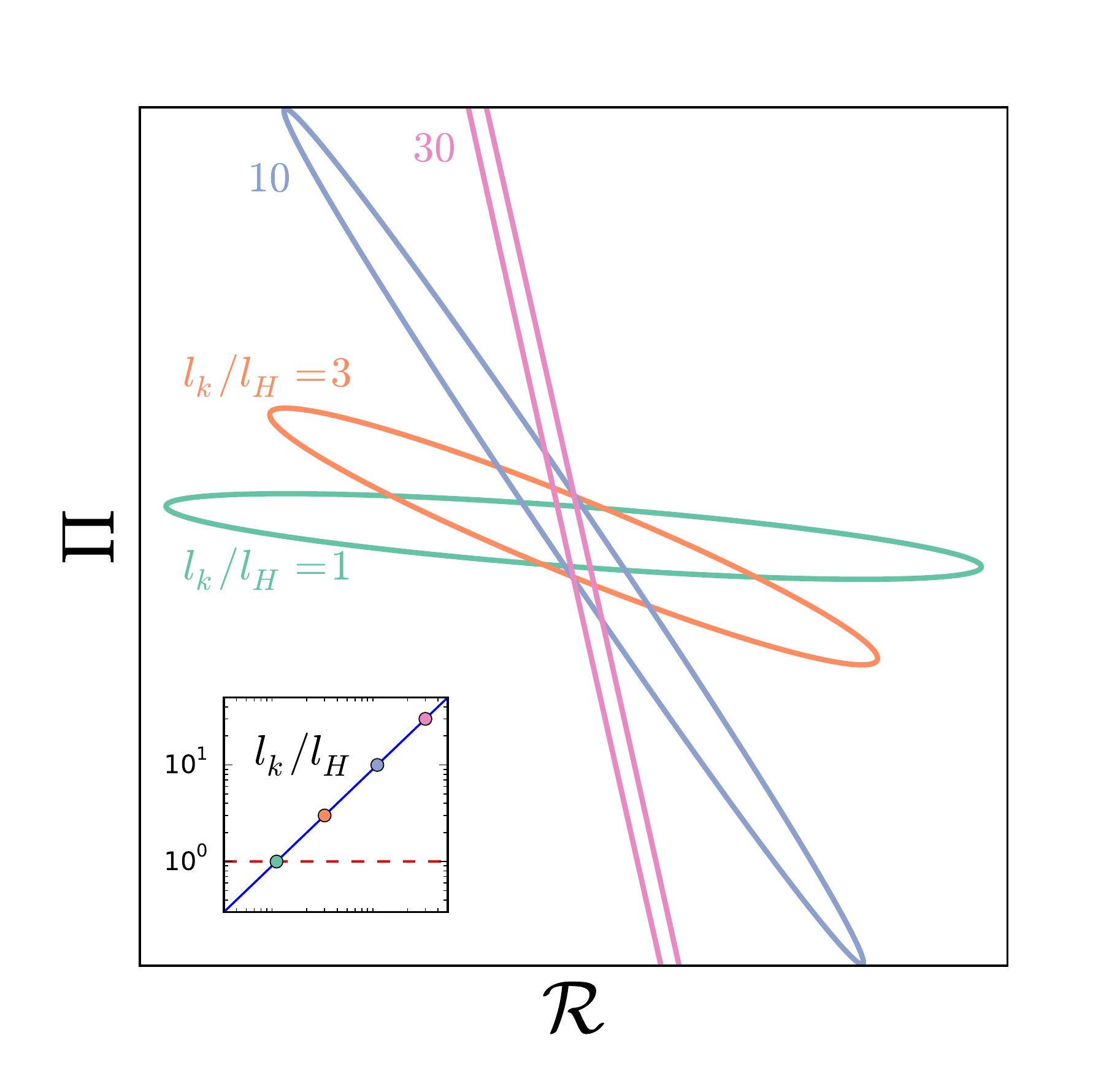} 
\caption{{\it Left:} The growing mode ($\mathcal{R}_{\rm grow}^{\rm SR}$, red) and decaying mode ($\mathcal{R}_{\rm dec}^{\rm SR}$, blue) characterizing the quantum state of the primordial perturbations in slow-roll inflation, as a function of the number of e-foldings of expansion since horizon exit, $N$. As a mode exits the horizon (at $N = 0$), the growing mode approaches a constant, while the decaying mode tends to zero. We have normalized the modes such that $\mathcal{R}_{\rm grow}^{\rm SR} \to 1$ in the super-horizon limit. {\it Right:} The rapid suppression (proportional to $\sim e^{-3N}$, where $N$ is the number of e-foldings since horizon exit) of the decaying mode relative to the growing mode leads to a highly squeezed state, characterized by a narrow Wigner function. We show (in arbitrary units) Wigner contours of constant $\chi^2 \equiv -2\ln (\pi W(\mathcal{R}, \Pi))$ for various values of $\ell_k/\ell_H = e^N$, the ratio of the mode wavelength to the Hubble scale (see inset). The highly squeezed super-horizon state behaves classically in the way discussed in the text.}
\label{fig:SR}
\end{figure*}

Let us now consider the evolution of the mode functions during inflation.
We see from Eq.~(\ref{eq:sols SR norm}) that while the mode is inside the horizon at early times, $| k \, \tau| \gg 1$, $\mathcal{R}_1(\tau)$ and $\mathcal{R}_2(\tau)$ oscillate, with a slowly varying (compared to the time scale of oscillations) and equal amplitude.
After horizon exit however, $|k \, \tau| \ll 1$,
the growing mode approaches a non-zero constant, $\mathcal{R}_1 \to \text{const.}$, while the decaying mode goes to zero.
This behavior is illustrated in the left panel of
Figure \ref{fig:SR},
which shows the normalized modes $a_\mathcal{R}^{-1} \, {\mathcal{R}}_1(\tau) = \mathcal{R}_{\rm grow}^{\rm SR}(\tau)$ and $a_\mathcal{R}^{-1} \, {\mathcal{R}}_2(\tau) = \mathcal{R}_{\rm dec}^{\rm SR}(\tau)$
as a function of the number of e-foldings since horizon exit,
\beq
e^{N} \equiv \frac{a(\tau)}{a(\tau_*)}= (-k \, \tau)^{-1} = \frac{l_k}{l_H},
\eeq
where $\tau_*$ is the time at which the mode $\k$ exits the horizon, $l_k \equiv k^{-1}$ is the comoving length scale of the mode and $l_H = (a \, H_I)^{-1}$ is the comoving Hubble length scale.
The quantity $e^N$ thus also gives the ratio of the wavelength of a mode $k$ to the Hubble scale.
At $|k\, \tau| \ll 1$, the decaying mode decays like $\mathcal{R}_2 \propto (-k \tau)^3$ so that the decaying mode is extremely rapidy suppressed relative to the growing mode,
\beq
\mathcal{R}_1(\tau) \gg \mathcal{R}_2(\tau) \quad \text{for} \quad |k \, \tau| \ll 1 \, (N > 0)
\eeq

The resulting state is a squeezed state, where $\mathcal{R}$ and $\Pi$ are highly correlated.
Indeed, Eq.~(\ref{eq:covWig}) shows that the correlation coefficient $\left|\text{Re}(\langle \hat{\mathcal{R}} \, \hat{\Pi} \rangle)/\sqrt{\langle \hat{\mathcal{R}}^2 \rangle \langle \hat{\Pi}^2 \rangle} \right| \to 1$.
The Wigner function (cf.~Eq.~(\ref{eq:covWig})) is thus stretched in the correlated direction and squeezed in the orthogonal direction, obtaining a cigarillo-like shape, as shown in the right panel of Figure \ref{fig:SR}.
Specifically, if we define the part of the momentum that is fully correlated with $\mathcal{R}$ as,
\beq
\label{eq:picl1}
\Pi_{\rm cl}(\mathcal{R}) \equiv
\frac{\text{Re}(\langle \hat{\mathcal{R}} \, \hat{\Pi}\rangle)}{\langle \hat{\mathcal{R}}^2 \rangle} \, \mathcal{R}
\approx \frac{z^2(\tau) \, \mathcal{R}_1'(\tau)}{\mathcal{R}_1(\tau)} \, \mathcal{R},
\eeq
the Wigner function approaches,
\beq
W(\mathcal{R}, \Pi) \to
\frac{1}{\sqrt{2 \pi \, {\bf C}_{\mathcal{R} \mathcal{R}}}} \,
\text{Exp}\left\{ -\frac{1}{2} \, ({\bf C}_{\mathcal{R} \mathcal{R}})^{-1} \, \mathcal{R}^2  \right\} \, \delta^{(D)}(\Pi - \Pi_{\rm cl}(\mathcal{R})),
\eeq
closely approximating a distribution along a single direction in phase-space.
A common alternative description of the squeezed Wigner function is in terms of a squeezing factor $r$, quantifying how stretched the Wigner function is, and a squeezing angle $\theta$, quantifying the direction in which it is stretched (both appearing in the singular value decomposition of the phase-space evolution matrix), see e.g.~\cite{Polarski:1995jg,Martin:2015qta}.
However, in this paper we choose to describe the state in terms of its growing and decaying modes and will not use those squeezing parameters.\footnote{We note that Eq.~(\ref{eq:picl1}) technically does not describe the squeezed direction if the state is squeezed exactly along the $\Pi$-axis (which is not the case during inflation, but is a situation of interest in the late Universe). In that case, the squeezed direction in phase space is defined by $\mathcal{R} = 0$. An advantage of the description in terms of the standard squeezing parameters is that $r$ is invariant under rotations in phase space and that it can describe squeezing in any direction, including along the $\Pi$-axis (or the $\mathcal{R}$-axis).}

\subsection{Classical behavior due to squeezing}
\label{subsec:squeezing2classical}

The resulting super-horizon squeezed state can be considered classical in the sense that $\hat{\mathcal{R}}$ and $\hat{\Pi}$ effectively commute in the limit of vanishing decaying mode.
From Eq.~(\ref{eq:evol zeta pi}),
\bea
\label{eq:zetapi squeezed}
\hat{\mathcal{R}}(\tau) &=& \sqrt{2} \, \mathcal{R}_2(\tau) \, \hat{x} - \sqrt{2} \, \mathcal{R}_1(\tau) \, \hat{p}
\approx - \sqrt{2} \, \mathcal{R}_1(\tau) \, \hat{p} \nonumber \equiv \hat{\mathcal{R}}_{\rm cl}(\tau) \\
\hat{\Pi}(\tau) &=& \sqrt{2} \, z^2(\tau) \, \mathcal{R}_2'(\tau) \, \hat{x} - \sqrt{2} \, z^2(\tau) \, \mathcal{R}_1'(\tau) \, \hat{p}  \approx - \sqrt{2} \, z^2(\tau) \, \mathcal{R}_1'(\tau) \, \hat{p} \equiv \hat{\Pi}_{\rm cl}(\tau),
\eea
with $[ \hat{\mathcal{R}}_{\rm cl}, \hat{\Pi}_{\rm cl} ] = 0$ (note that, despite the ``cl'' subscript, $\hat{\mathcal{R}}_{\rm cl}$ and $\hat{\Pi}_{\rm cl}$ are still operators).
The commuting components of the operators, $\hat{\mathcal{R}}_{\rm cl}$ and $\hat{\Pi}_{\rm cl}$, are given by the growing mode $\mathcal{R}_1$
and the non-commuting remainders are suppressed by the decaying mode\footnote{This ``hiding'' of the commutator of $\hat{\mathcal{R}}$ and $\hat{\Pi}$ is a general result for squeezed states
(i.e.~it is not specific to the squeezed state during slow-roll inflation). A slight subtlety is that, for a general solution of the form Eq.~(\ref{eq:evol zeta pi}), one always has the freedom to redefine $\mathcal{R}_1$ and $\mathcal{R}_2$ by a rotation without changing the solution (see discussion at the end of Section \ref{subsec:sri}). In the description of the classical limit in this Section, it is thus implicit that one mode (here $\mathcal{R}_1$) is chosen to be the dominant/growing mode, and the other (here $\mathcal{R}_2$) the minimal/decaying mode, so that (the norm in phase space of) $\mathcal{R}_1$ is maximally dominant over (the norm of) $\mathcal{R}_2$.} $\mathcal{R}_2$.
In the sense that the quantum nature of the perturbations is captured by the lack of commutation between operators, we may loosely consider the operators $\hat{\mathcal{R}}_{\rm cl}$ and $\hat{\Pi}_{\rm cl}$
as the ``classical''  components (hence the subscript).
Note that these classical components correspond to the stretched/correlated direction of the Wigner function in the previous subsection, with $\hat{\Pi}_{\rm cl} \approx \Pi_{\rm cl}(\hat{\mathcal{R}})$.

Suppression of the non-commuting component of the perturbations leads to classical behavior in the concrete sense that
expectation values become extremely well approximated by expectation values of a classical stochastic distribution.
As discussed at the end of Section \ref{subsubsec:wigner}, if we treat $\hat{\mathcal{R}}$ and $\hat{\Pi}$ (or equivalently $\hat{x}$ and $\hat{p}$ in Eq.~(\ref{eq:zetapi squeezed})) as variables drawn from a stochastic ensemble (instead of as operators) that evolve classically with probability distribution equal to the Wigner function, this does {\it not in general} reproduce the proper quantum statistics of the system. However, in the squeezed limit, the classical description in terms of the Wigner function {\it does} approach the full quantum expectation value,
\beq
\label{eq:class exp}
\langle A(\hat{\mathcal{R}}, \hat{\Pi}) \rangle \approx \langle A(\hat{\mathcal{R}}, {\Pi}_{\rm cl}(\hat{\mathcal{R}})) \rangle
= \int d\mathcal{R} \, |\psi(\mathcal{R})|^2 \, A(\mathcal{R}, \Pi_{\rm cl}(\mathcal{R}))
= \int d\mathcal{R} \, d\Pi \, W(\mathcal{R}, \Pi) \, A(\mathcal{R}, \Pi),
\eeq
where we have used that marginalizing the Wigner function over one direction in phase space gives the probability distribution (i.e.~the square of the wave function) of the other direction, and specifically
$|\psi(\mathcal{R})|^2 = \int  d\Pi \, W(\mathcal{R}, \Pi)$.
For instance, the (absolute value of the) ``quantum'', imaginary part of
the cross-correlation between $\hat{\mathcal{R}}$ and $\hat{\Pi}$, $\text{Im}(\langle \hat{\mathcal{R} \, \hat{\Pi}}\rangle) = \iu/2$
is now negligible compared to the real part, $|\text{Re}(\langle \hat{\mathcal{R}} \, \hat{\Pi} \rangle)| \gg 1$, which {\it is} captured by the classical description\footnote{Technically, one could of course
consider combinations such as $\hat{\Pi} - \Pi_{\rm cl}(\hat{\mathcal{R}})$, i.e.~subtracting out the growing mode component that commutes with $\hat{\mathcal{R}}$. One is then explicitly probing the squeezed direction in phase space. Expectation values involving such quantities
are still not well described by the classical description, in the sense that the difference between the classical prediction and the quantum expectation value is large compared to the latter. For instance, $\langle \hat{\mathcal{R}} \, (\hat{\Pi} - \Pi_{\rm cl}(\hat{\mathcal{R}})) \rangle = \iu/2$, while the classical description would give zero. As we will see, however, the non-commuting/decaying mode component of the perturbations actually becomes very small compared to any reasonable observational uncertainties, so even for the above expectation value where the ``quantum component'' dominates, this component would still be unmeasurably small.
}.
Moreover, the classical treatment now reproduces expectation values of higher order statistics to good approximation (see also \cite{martinvennin16}).
For instance, the observable 4-point function corresponding to the Hermitian operator $\tfrac{1}{2} (\hat{\mathcal{R}}^2 \, \hat{\Pi}^2 + \hat{\Pi}^2 \, \hat{\mathcal{R}}^2)$ has the quantum expectation value,
\beq
\frac{1}{2} \, \langle \hat{\mathcal{R}}^2 \, \hat{\Pi}^2 + \hat{\Pi}^2 \, \hat{\mathcal{R}}^2 \rangle = 3 \langle \hat{\mathcal{R}}^2 \rangle \, \langle \hat{\Pi}^2 \rangle  - 1 = 3 \left(\mathcal{R}_1^2(\tau) + \mathcal{R}_2^2(\tau) \right) \, z^2(\tau) \, \left( \mathcal{R}_1^{' \, 2}(\tau) + \mathcal{R}_2^{' \, 2}(\tau) \right) - 1,
\eeq
whereas treating the Wigner function as a classical distribution instead gives,
\beq
\frac{1}{2} \, \langle {\mathcal{R}}^2 \, {\Pi}^2 + {\Pi}^2 \, {\mathcal{R}}^2 \rangle_{\rm cl}
= \langle \mathcal{R}^2 \, \Pi^2 \rangle_{\rm cl}
= 3 \langle \hat{\mathcal{R}}^2 \rangle \, \langle \hat{\Pi}^2 \rangle  - \frac{1}{2}.
\eeq
For the initial vacuum state, we have $\langle \hat{\mathcal{R}}^2 \rangle \, \langle \hat{\pi}^2 \rangle = 1/4$
so that the two expectation values have an order unity difference. Indeed, the quantum expectation value is negative (equal to $-1/4$), while the classical estimate is still positive (as it should be).
In the squeezed limit, on the other hand, $\langle \hat{\mathcal{R}}^2 \rangle \, \langle \hat{\Pi}^2 \rangle \approx \mathcal{R}_1^2 \, z^2 \, \mathcal{R}_1^{' \, 2} \gg 1/4$ so that the relative difference becomes smaller the more squeezed the state is.
We note that the difference of $1/2$ between the two expectation values traces back to products of two growing mode and two decaying mode contributions, thus confirming that the classical limit corresponds to $\mathcal{R}_2 \ll \mathcal{R}_1$.




It is in the above sense that squeezing achieves a quantum-to-classical transition.
We stress however that, in another sense, the state is still very ``quantum'': we are still describing the perturbations by a pure quantum state in a coherent superposition of a (observably) large range of $\mathcal{R}$ values. This is very different from the coherent states (having small quantum spread in both directions in phase space) that are conventionally considered ``classical''.
The quantum-to-classical transition by squeezing has thus been called
{\it decoherence without decoherence} \cite{Polarski:1995jg}.


\subsection{Searching for a remaining ``quantum signature''}
\label{subsec:searching}

The decaying mode, which quantifies the non-commuting component of the phase-space operators, thus decribes deviations from the classical limit in the specific sense discussed above. Therefore,
{\bf we will use the contribution from the decaying mode to quantify the quantumness of the inflationary perturbations}, and loosely define the ``quantum component'' (cf.~Eq.~(\ref{eq:zetapi squeezed})),
\bea
\label{eq:zetapi quantum}
\delta \hat{\mathcal{R}}_{\rm qu}(\tau) &\equiv& \sqrt{2} \, \mathcal{R}_2(\tau) \, \hat{x} \\
\delta \hat{\Pi}_{\rm qu}(\tau) &\equiv& \sqrt{2} \, z^2(\tau) \, \mathcal{R}_2'(\tau) \, \hat{x} \approx \hat{\Pi}(\tau) - \Pi_{\rm cl}(\hat{\mathcal{R}}(\tau)).
\eea
Specifically, we will consider the perturbations in the late Universe, when observations are made, and will quantify the amplitude of the late-Universe decaying and growing modes.
The end goal is to compare the effect of the decaying mode on $\hat{\mathcal{R}}$ and $\hat{\Pi}$ to observational errors (at an order of magnitude level) and to see if there are scenarios where this ``quantum component'' may be large enough to be in principle observable.
In the standard inflationary scenario, we will of course recover the well known result that
the state is extremely squeezed by the end of inflation and that moreover, the state gets squeezed even further in the post-inflationary Universe during the many e-foldings of expansion during which the mode is super-horizon.
The result is that in the late-Universe
the decaying mode is suppressed by $\sim 115$ orders of magnitude, making it almost comically unobservable. We will then ask whether this conclusion can be altered in non-standard inflationary scenarios.

\begin{figure*}[]
\centering
\includegraphics[width=0.7\textwidth]{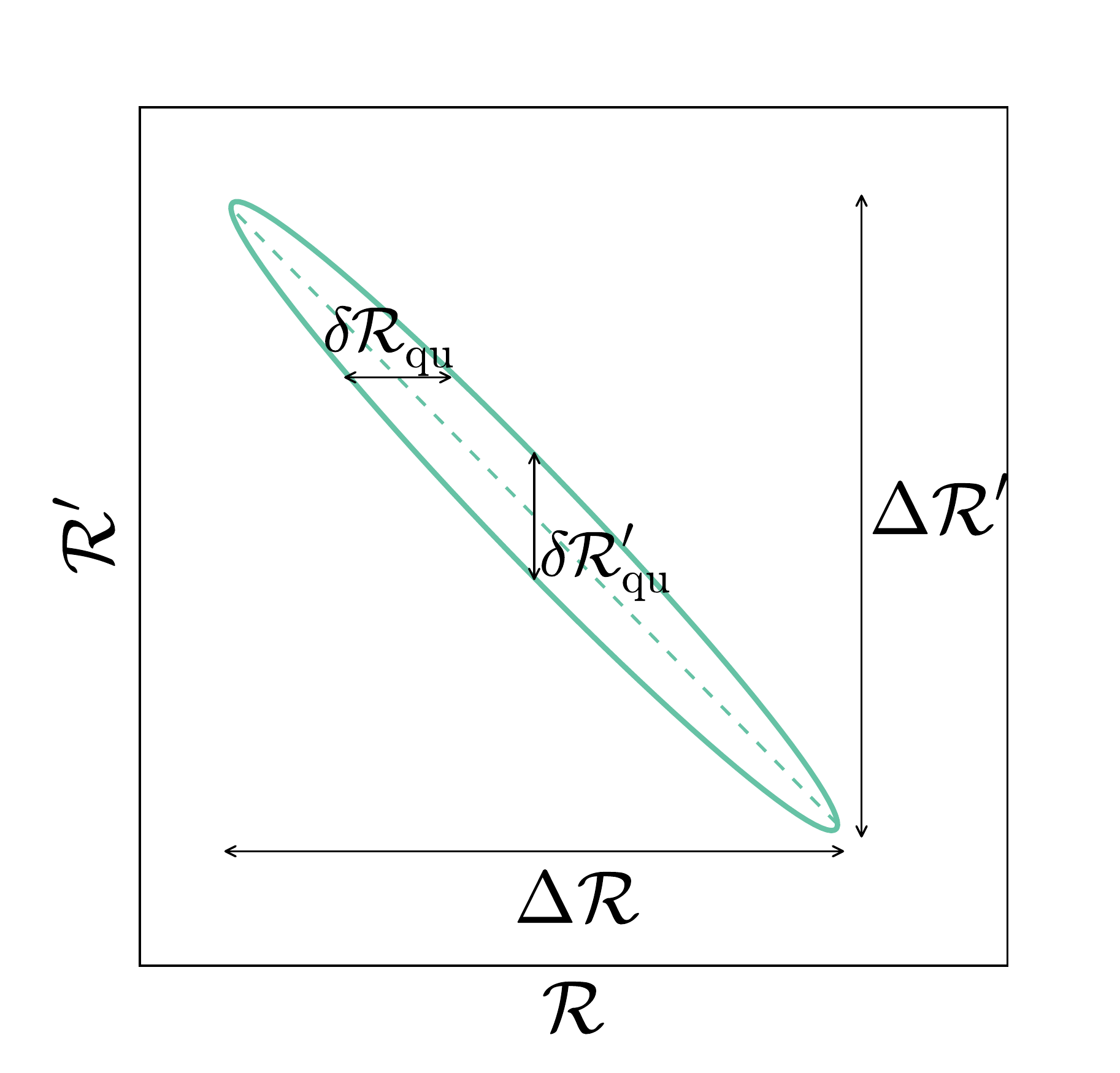} 
\caption{A Wigner ellipse ($\chi^2 = 1$, see text) in the squeezed limit, in terms of $\mathcal{R}$ and $\mathcal{R}'$. 
The extent in the $\mathcal{R}$ and $\mathcal{R}'$ directions gives the root-mean-square values of these operators (just like $\chi^2 = 1$ contours of a true phase-space distribution would).
In the squeezed limit, the extended direction of the ellipse (dashed line) is described by the growing mode and the growing mode components of the operators $\hat{\mathcal{R}}$ and $\hat{\mathcal{R}}'$ commute. We thus colloquially refer to the growing mode contribution as the {\it classical} component.
The decaying mode describes the non-zero width of the ellipse.
Since it is the inclusion of the decaying mode that causes the non-commutation of $\hat{\mathcal{R}}$ and $\hat{\mathcal{R}}'$,
we will loosely refer to the decaying mode as the ``quantum'' contribution.
With this convention, the ``quantum'' spread in $\mathcal{R}$ and $\mathcal{R}'$, $\delta \mathcal{R}_{\rm qu}$ and $\delta \mathcal{R}_{\rm qu}'$ are as indicated in the Figure and become small relative to the ``classical'' spread as the decaying mode gets more and more suppressed and the Wigner ellipse narrower.
We stress that the primordial state under consideration is fully quantum mechanical even in the squeezed limit.
We here use the labels ``quantum'' and ``classical'' only in the specific sense explained above and in the text.
}
\label{fig:wigner illus}
\end{figure*}

To connect more directly to late-Universe observations, we will in this paper directly describe the phase space in terms of $\hat{\mathcal{R}}$ and its derivative $\hat{\mathcal{R}}'$ (the latter in place of the canonical momentum $\hat{\Pi} = z^2 \hat{\mathcal{R}}'$), and define the ``quantum'' and ``classical'' components of $\hat{\mathcal{R}}'$ analogously to Eqs (\ref{eq:zetapi squeezed}) and (\ref{eq:zetapi quantum}).
We illustrate the extent of the quantum (and classical) components in the Wigner ellipse ($\chi^2 = 1$) in Figure \ref{fig:wigner illus}.
The stretched direction describes the strong correlation between $\hat{\mathcal{R}}$ and $\hat{\mathcal{R}}'$ and corresponds to the ``classical'' direction determined by the growing mode.
The extent of the contour in the $\mathcal{R}$ direction gives the rms fluctuation $\Delta \mathcal{R} \equiv \sqrt{\langle \hat{\mathcal{R}}^2 \rangle} \approx \sqrt{\langle \hat{\mathcal{R}}_{\rm cl}^2 \rangle}$ and the extent in the $\mathcal{R}'$ direction gives the rms fluctuation $\Delta \mathcal{R}' \equiv \sqrt{\langle \hat{\mathcal{R}}^{' \, 2} \rangle} \approx \sqrt{\langle \hat{\mathcal{R}}_{\rm cl}^{' \, 2} \rangle}$.
The ``quantum'' contribution is responsible for the non-zero width of the squeezed direction, giving subdominant contributions to the variance in the $\mathcal{R}$ and $\mathcal{R}'$ directions.

\vskip 7pt

We caution here that merely detecting the decaying mode (or quantum) component does not in itself consitute a ``smoking gun'' of the quantum nature of the fluctuations. For instance, if we detect it in the power spectrum of $\mathcal{R}$, that same observation can still be decribed in terms of classical, stochastic curvature perturbations occupying both the growing and decaying mode.
However, there will be other observables that can {\it not} be reproduced in a classical description (e.g.~4-point functions) and if the decaying mode is detectably large, those observables may truly distinguish between the quantum and classical descriptions.
In this paper, our first focus is on the first step above, i.e.~can we in principle detect the decaying mode in, say, the power or cross-spectra of $\mathcal{R}$ and $\mathcal{R}'$? If we find a scenario where the answer is yes, it then makes sense to ask what specific, more complicated observations might provide a true smoking gun of the quantum nature.

An important caveat to the analysis in this paper is that we will ignore decoherence \cite{decoherencebook} and treat the quantum state of each mode ${\bf k}$ as a pure state even in the late Universe.
In reality, interactions of a mode with other modes (due to non-linearity in the action, e.g.~\cite{nelson16}), as well as interactions with other degrees of freedom (e.g.~\cite{proprigo07}), will entangle each mode with its {\it environment}.
If we now consider the system constituded by a single mode, it is described by a mixed state, not a pure state.
Thus, the quantum superposition between different values of $\mathcal{R}$ becomes incoherent and effectively, the mode is measured by its environment.
We consider our treatment of each mode as a pure state an idealized scenario and a useful starting point for an understanding of the perturbations in the presence of decoherence.
Moreover, since generally decoherence plays a key role in explaining the quantum-to-classical transition,
decoherence is expected to make the state less ``quantum''.
Thus, if in the pure state approximation, we cannot find an observable quantum signal, this conclusion is not likely to change with the inclusion of decoherence.
We will comment more on the role of decoherence in Section \ref{subsec:dec}.

\section{Evolution of the quantum state after slow-roll inflation}
\label{sec:SR2RD}

We will now study the evolution of the growing and decaying modes from inflation into the late Universe, where observations are made, in the standard scenario of single-field slow-roll inflation.
For simplicity, we will model the entire post-inflationary phase as a radiation-dominated (RD) Universe.
This will be sufficient for our purposes of deriving physical insights into the late-Universe quantum signature (or lack thereof).
If we do find a potentially observable signal, we may then consider a less crude description of the late Universe and
include the effects of pressureless matter, neutrinos and dark energy.

\begin{figure*}[]
\centering
\includegraphics[width=0.75\textwidth]{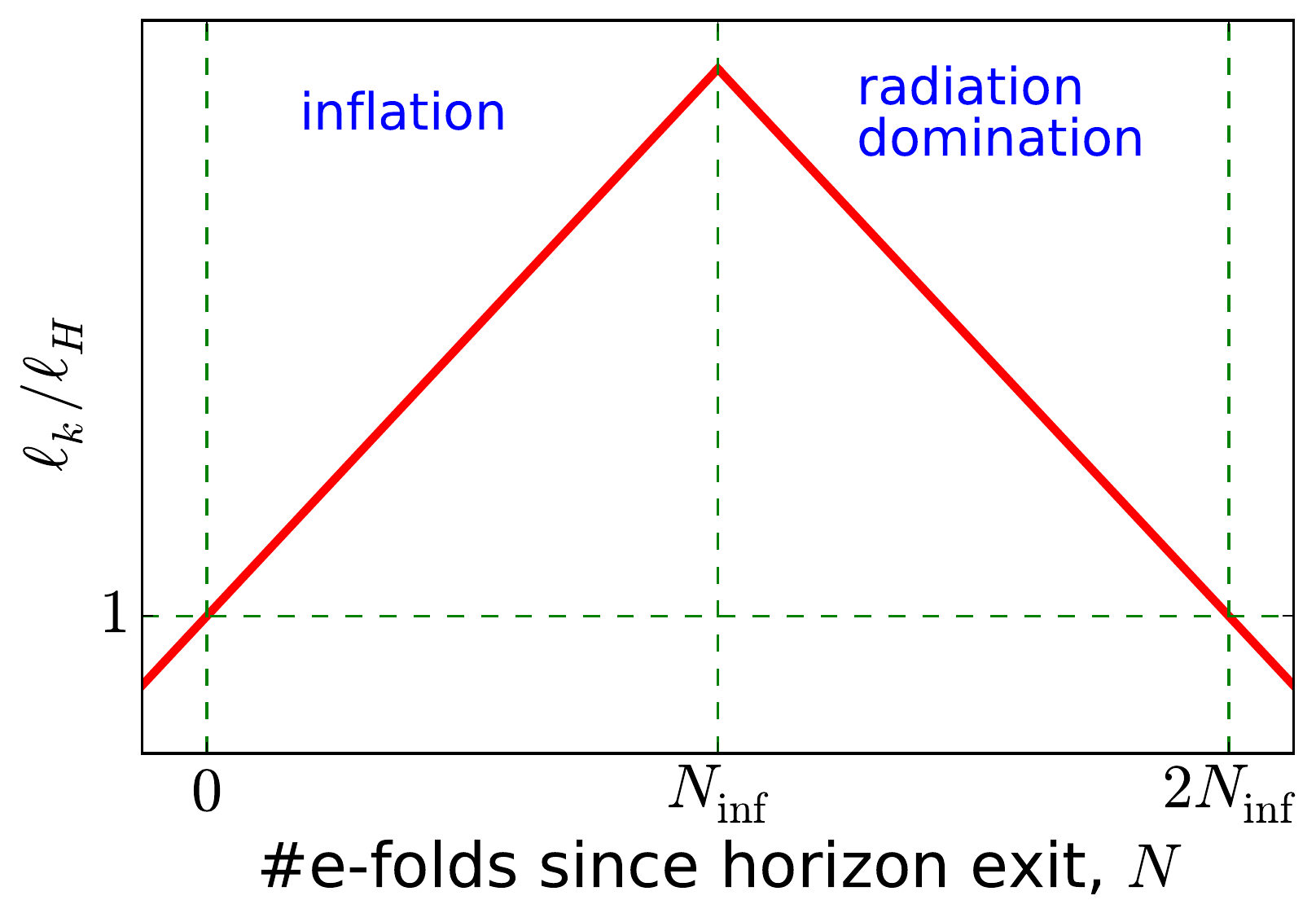} 
\caption{Ratio of the wavelength of a mode with wave number $k$ to the Hubble scale, as a function of the number of e-foldings of expansion since horizon exit. There are $N_*$ e-foldings between horizon exit and the end of inflation. We then assume an instantaneous transition to a radiation dominated hot big bang phase. In a simplified Universe where the Universe remains radiation dominated (and the number of radiation degrees of freedom is constant), it takes exactly another $N_*$ e-foldings before the mode $k$ re-enters the horizon.}
\label{fig:lk/lH}
\end{figure*}

We will consider an instantaneous transition
where inflation ends at some time $\tau = - \tau_e < 0$. After the transition, the Universe is taken to be radiation dominated and conformal time continues from $\tau = \tau_e > 0$. With this convention, the scale factor (which is continuous through the transition) in the radiation-dominated epoch is,
\beq
a(\tau) = \frac{\tau}{H_e \, \tau_e^2},
\eeq
where $H_e$ is the Hubble parameter (also assumed continuous) at the time of the transition $\tau_e$. In our case, we consider the de Sitter limit where the Hubble parameter during inflation is constant, $H(\tau) = H_I$, so that $H_e = H_I$.
The ratio of the mode scale to the Hubble scale during RD decreases with time as,
\beq
\frac{\ell_k}{\ell_H} = \frac{a(\tau) \, H(\tau)}{k} = (k \, \tau)^{-1} = e^{2 N_* - N},
\eeq
where $N_* = N(\tau_e)$ is the number of e-foldings between horizon exit and the end of inflation.
Since we assume RD in the late Universe, it thus takes exactly another $N_*$ e-foldings after inflation for the mode to enter the horizon again. By this time, the mode has undergone a total of $2 N_*$ e-foldings of expansion since horizon exit.
We illustrate this in Figure \ref{fig:lk/lH}.

During radiation domination, the perturbations are described by the action of a perfect fluid \cite{mukhbrandfeld92} already given in Eq.~(\ref{eq:action}),
where now,
\beq
\epsilon^{\rm RD} = \tfrac{3}{2} \, (1 + w^{\rm RD}) = 2,
\eeq
and the sound speed is,
\beq
c_s^{\rm RD} = \sqrt{\frac{1}{3}}.
\eeq
The equation of motion is given by Eq.~(\ref{eq:EoM}), which has the two independent solutions given by the spherical Bessel functions of the first and second kind respectively,
\bea
\label{eq:modesRD}
\mathcal{R}_{\rm grow}^{\rm RD}(\tau) &\equiv& \frac{\sin\left(\sqrt{\tfrac{1}{3}} \, k \, \tau \right)}{\sqrt{\tfrac{1}{3}} \, k \, \tau} \nonumber \\
\mathcal{R}_{\rm dec}^{\rm RD}(\tau) &\equiv& \frac{\cos\left( \sqrt{\tfrac{1}{3}} \, k \, \tau \right)}{\sqrt{\tfrac{1}{3}} \, k \, \tau}.
\eea
We have normalized the solutions such that they are of the same order at the time of horizon entry, $k \, \tau \sim 1$.

\vskip 7pt

To quantify the late-Universe quantum signature, we will expand $\hat{\mathcal{R}}$ and $\hat{\Pi}$ (or $\hat{\mathcal{R}}'$) in terms of the above RD growing and decaying modes, in the same form as Eq.~(\ref{eq:evol zeta pi}).
To do this, we first compute
the evolution of the inflationary modes $\mathcal{R}_{\rm grow}^{\rm SR}(\tau)$ and $\mathcal{R}_{\rm dec}^{\rm SR}(\tau)$ into the radiation dominated epoch, specifically constructing the
linear combinations of $\mathcal{R}_{\rm grow}^{\rm RD}(\tau)$ and $\mathcal{R}_{\rm dec}^{\rm RD}(\tau)$ that $\mathcal{R}_1(\tau)$ and $\mathcal{R}_2(\tau)$ evolve into.

\vskip 7pt

We thus need to evolve the initial modes through the reheating transition at $\tau_e$.
To do this, we assume a simple toy model where the curvature perturbations are described by the perfect fluid action (\ref{eq:action}) at all times, including during the transition. The transition can then be seen as a simple change in functions $\epsilon(\tau) = \tfrac{3}{2} \, (1 + w(\tau))$ and $c_s(\tau)$ from (constant) values $(\epsilon, c_s) = (\epsilon_I, 1)$ to $(2, \sqrt{1/3})$, where $\epsilon_I \ll 1$ is the value during slow-roll inflation.
This is a very simplified decription of reheating (we refer to \cite{deruellemukhanov95,durvern02,mukhbook} for more general discussions of matching conditions between cosmic phases) and neglects any effect that entropy perturbations associated with the Universe having multiple components might have on $\mathcal{R}$.
In the above picture, the general equation of motion, Eq.~(\ref{eq:EoM}), applies throughout the transition.
In the limit where the transition is instantaneous, or at least occurs on a time scale much shorter than the Hubble time,
it then follows from the equation of motion that
$\mathcal{R}(\tau)$ and its conjugate momentum $\Pi(\tau) = z^2(\tau) \, \mathcal{R}'(\tau)$ are continuous, so that,
\bea
\label{eq:matching1}
\mathcal{R}_+ &=& \mathcal{R}_- \quad \text{(continuous)} \nonumber \\
\mathcal{R}'_+ &=& \frac{\left( 2 \epsilon/c_s^2 \right)_-}{\left( 2 \epsilon/c_s^2 \right)_+} \, \mathcal{R}_-'
= \frac{\epsilon_e}{6} \, \mathcal{R}_-' = \frac{\epsilon_*}{6} \, \mathcal{R}_-' ,
\eea
where $-$ and $+$ indicate the values before and after the transition, respectively, and $\epsilon_e$ is the inflationary slow-roll parameter at the time of the transition.
In the present scenario, we take $\epsilon$ to be constant during inflation, so that in particular $\epsilon_e = \epsilon_*$, the value at the time of horizon exit.
We will in the next Section consider the ultra-slow-roll scenario where $\epsilon$ evolves during inflation, so that $\epsilon_e \ne \epsilon_*$.
We see from the matching conditions in Eq.~(\ref{eq:matching1}) that the derivative $\mathcal{R}'(\tau)$ is discontinuous through the transition.

We can now apply the matching conditions above to the inflationary growing and decaying modes given in Eq.~(\ref{eq:sols SR}) (and Eq.~(\ref{eq:sols SR norm})),
\bea
\mathcal{R}_1(\tau) &=& a_\mathcal{R} \, \mathcal{R}_{\rm grow}^{\rm SR}(\tau) =  - a_\mathcal{R} \, \sqrt{\frac{\pi}{2}} \, (-k \, \tau)^{3/2} \, Y_{3/2}(-k \, \tau) \quad  \text{\bf (during SR inflation)} \nonumber \\
\mathcal{R}_2(\tau) &=& a_\mathcal{R} \, \mathcal{R}_{\rm dec}^{\rm SR}(\tau) = - a_\mathcal{R} \, \sqrt{\frac{\pi}{2}} \, (-k \, \tau)^{3/2} \, J_{3/2}(- k \, \tau) \quad \, \text{\bf (during SR inflation)} \nonumber,
\eea
so that just after the transition (i.e.~at $\tau = \tau_e$), we can write the two phase-space vectors in matrix form as,
\beq
{\bf E}_{1,2} \equiv \begin{pmatrix} \, \mathcal{R}_1(\tau_e)  & \, \mathcal{R}_2(\tau_e) \\ \mathcal{R}_1'(\tau_e)  & \, \, \mathcal{R}_2'(\tau_e) \end{pmatrix}
= a_\mathcal{R} \, \begin{pmatrix} \, \mathcal{R}_{\rm grow}^{\rm SR}(-\tau_e)  & \, \mathcal{R}_{\rm dec}^{\rm SR}(-\tau_e) \\ \epsilon_*/6 \, (\mathcal{R}_{\rm grow}^{\rm SR})'(-\tau_e)  & \, \, \epsilon_*/6 \, (\mathcal{R}_{\rm dec}^{\rm SR})'(-\tau_e) \end{pmatrix}.
\eeq
To find how $\mathcal{R}_1(\tau)$ and $\mathcal{R}_2(\tau)$ continue to evolve during radiation domination, we express $\mathcal{R}_1(\tau)$ and $\mathcal{R}_2(\tau)$ at $\tau > \tau_e$ as linear combinations of the RD growing and decaying modes,
\bea
\mathcal{R}_1(\tau) &=& a \, \mathcal{R}_{\rm grow}^{\rm RD}(\tau) + b \, \mathcal{R}_{\rm dec}^{\rm RD}(\tau) \quad  \text{\bf (during RD)} \nonumber \\
\mathcal{R}_2(\tau) &=& c \, \mathcal{R}_{\rm grow}^{\rm RD}(\tau) + d \, \mathcal{R}_{\rm dec}^{\rm RD}(\tau) \quad  \text{\bf (during RD)}.
\eea
Finding the coefficients is now a simple linear algebra problem corresponding to applying a basis transformation at $\tau = \tau_e$ with tranformation matrix,
\beq
{\bf T} \equiv \begin{pmatrix} \, a & \, c \\ b & \, \,d \end{pmatrix} = \left({\bf E}_{\rm RD}\right)^{-1} \, {\bf E}_{1,2},
\eeq
with the ${\bf E}$ matrices containing the basis vectors as columns
and specifically,
\beq
{\bf E}_{\rm RD} = \begin{pmatrix} \, {\mathcal{R}}_{\rm grow}^{{\rm RD}}(\tau_e)  & \, {\mathcal{R}}_{\rm dec}^{{\rm RD}}(\tau_e) \\ \mathcal{R}_{\rm grow}^{{\rm RD}'}(\tau_e)  & \, \mathcal{R}_{\rm dec}^{{\rm RD}'}(\tau_e) \end{pmatrix}.
\eeq

We compute the transformation matrix ${\bf T}$ by expanding the mode vectors in terms of the small quantity $x_e = k \, \tau_e = e^{-N_*} \ll 1$,
giving to leading order,
\beq
\label{eq:zeta1RD}
a_\mathcal{R}^{-1} \, \mathcal{R}_1(\tau) =
\left(1
+ \mathcal{O}(x^2_e) \right) \,  \mathcal{R}_{\rm grow}^{\rm RD}(\tau) +
\left(-\frac{\sqrt{3}}{27} \, \left(1 - \tfrac{3}{2} \, \epsilon_* \right) \, x_e^3  + \mathcal{O}(x_e^5) \right) \, \mathcal{R}_{\rm dec}^{\rm RD}(\tau) \, \,\, \text{\bf (during RD)},
\eeq
and
\beq
\label{eq:zeta2RD}
a_\mathcal{R}^{-1} \, \mathcal{R}_2(\tau) =
\left( - \frac{1}{3} \, \left( 1 - \tfrac{1}{2} \, \epsilon_* \right) \, x_e^3 + \mathcal{O}(x_e^5) \right) \, \mathcal{R}_{\rm grow}^{\rm RD}(\tau)
+ \left(  - \frac{\epsilon_*}{6 \sqrt{3}} \, x_e^4 + \mathcal{O}(x_e^6) \right) \, \mathcal{R}_{\rm dec}^{\rm RD}(\tau) \, \,\, \text{\bf (during RD)}. \nonumber
\eeq

The inflationary growing mode $\mathcal{R}_1$ thus evolves into the RD growing mode, up to a tiny correction, that at $\tau_e$ is suppressed by $x_e^2$ (recall that, at $\tau_e$, $\mathcal{R}_{\rm dec}^{\rm RD} \sim x_e^{-1}$).
For this mode, $\mathcal{R}$ approaches a constant after horizon exit, remains constant during and after the transition to RD while the mode is super-horizon, and only starts to evolve again when it enters the horizon at late times.

The inflationary {\it decaying} mode, $\mathcal{R}_2$, decays as $(-k \, \tau)^3$ while the mode is super-horizon during slow-roll inflation.
Therefore, by the end of inflation, it is already suppressed by a factor $x_e^3 = e^{-3 N_*}$ relative to the growing mode.
For this reason,
most analyses of the primordial perturbations are solely concerned with the behavior of the growing mode.
However, even though it is indeed extremely suppressed in the standard scenario, we here consider the decaying mode more quantitatively into the RD regime.
Interestingly, $\mathcal{R}_2$ also predominantly evolves into the RD growing mode. While at $\tau_e$,
both contributions are of order $x_e^3$, the decaying mode contribution is additionally suppressed by the slow-roll parameter during inflation, $\epsilon_*$.
This is a direct consequence of the matching conditions, Eq.~(\ref{eq:matching1}), and in particular of the discontinuity in $\mathcal{R}'$ at $\tau_e$.
Thus, $\mathcal{R}_2$ becomes a constant, proportional to $\mathcal{R}_{\rm grow}^{\rm RD}$, during radiation domination. The amplitude of this mode is suppressed by the aforementioned factor $\sim e^{-3 N_*}$ relative to the mode $\mathcal{R}_1$.

\begin{figure*}[]
\centering
\includegraphics[width=0.32\textwidth,height=0.25\textwidth]{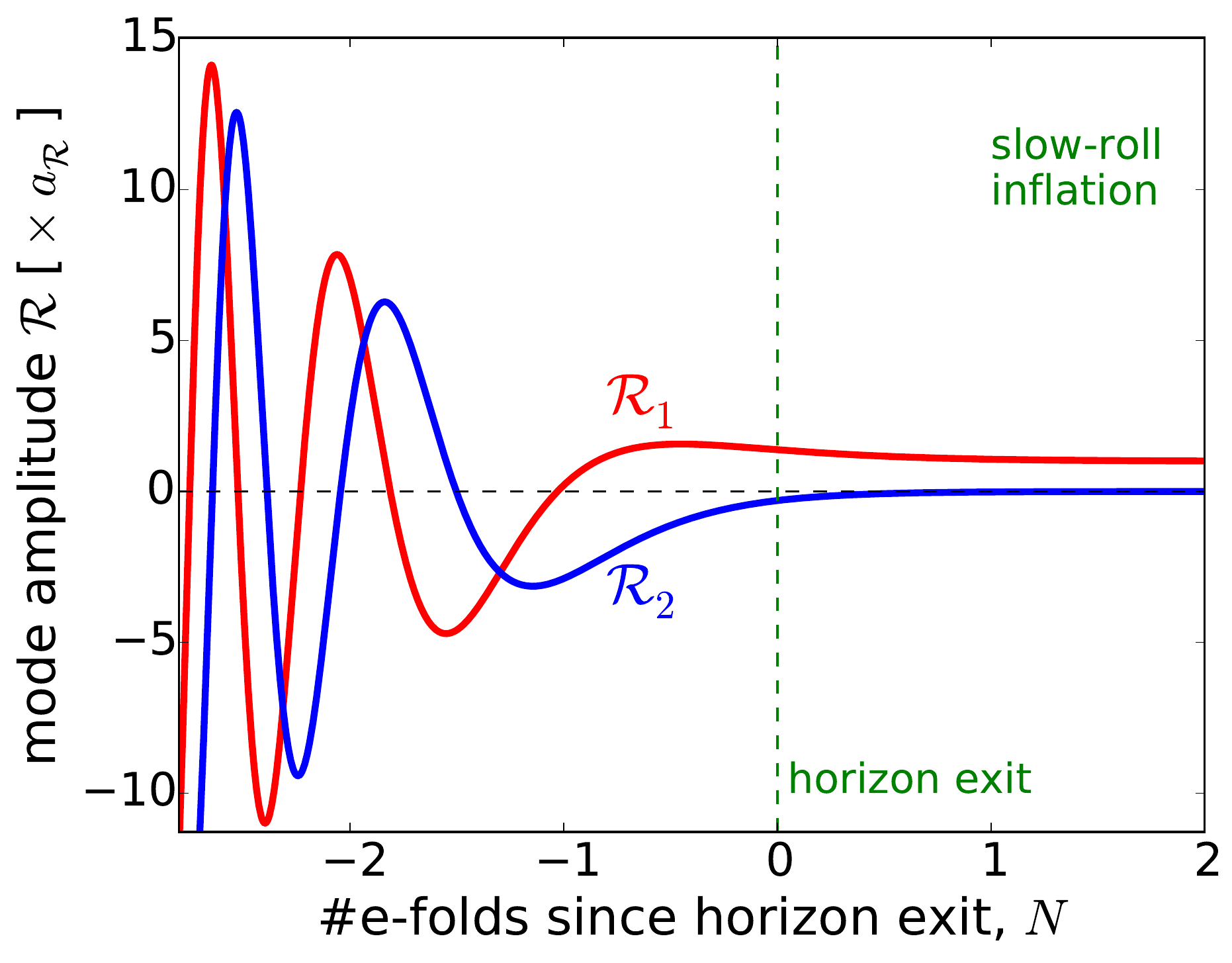} 
\includegraphics[width=0.32\textwidth,height=0.25\textwidth]{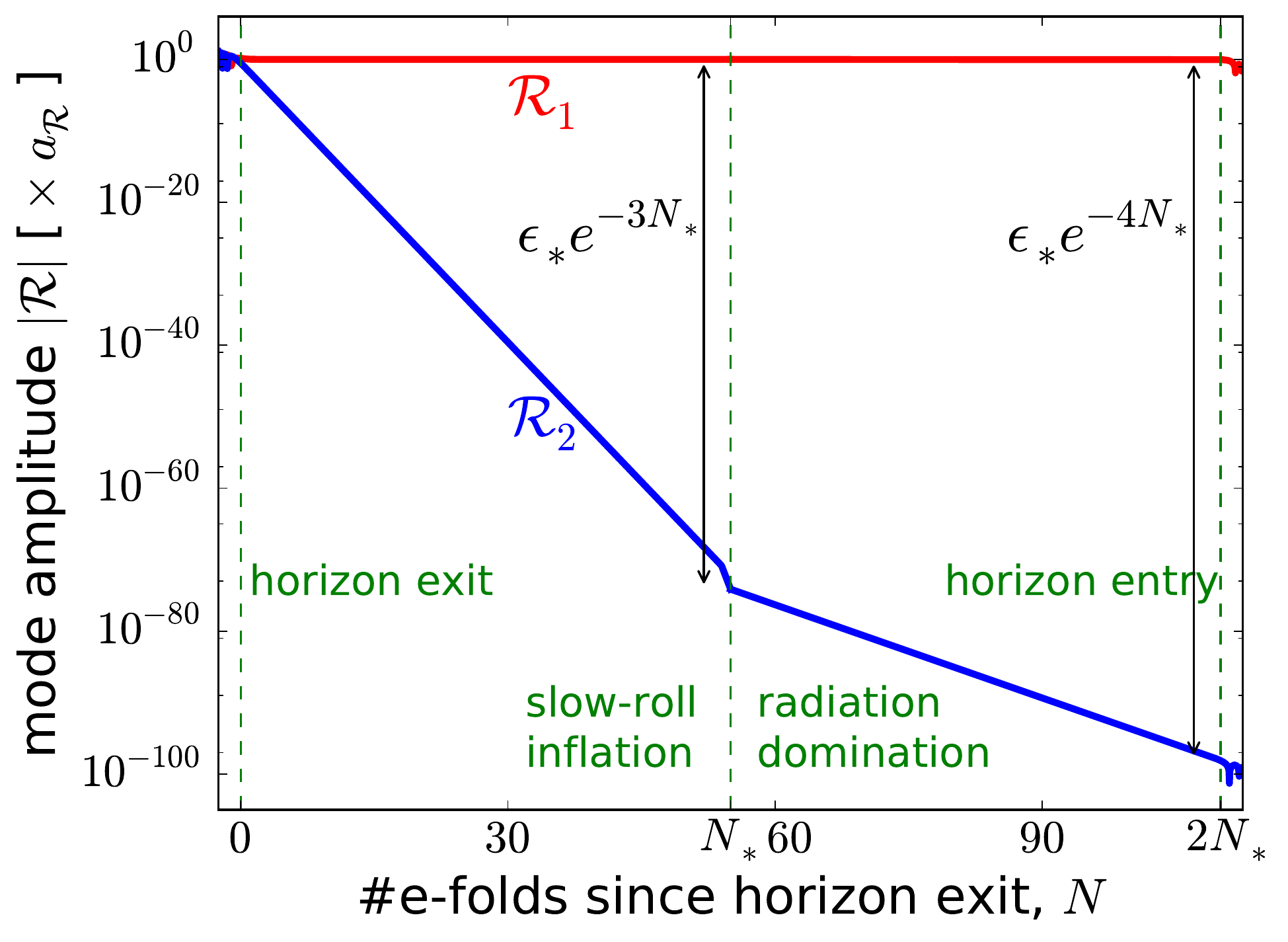} 
\includegraphics[width=0.32\textwidth,height=0.25\textwidth]{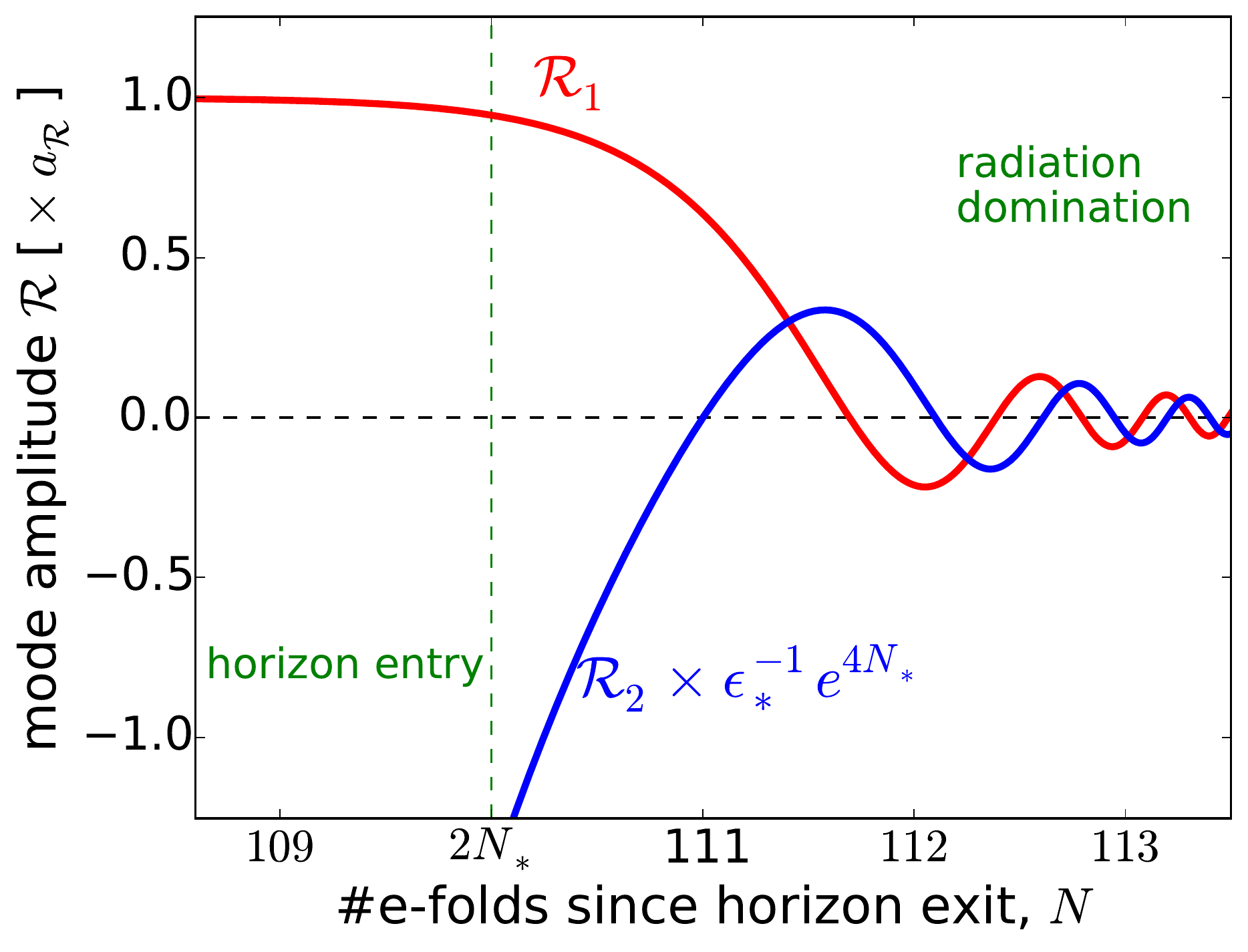} 
\caption{Evolution of the growing and decaying modes describing the primordial quantum state in a simplified version of the slow-roll scenario,
where the post-inflationary phase is radiation dominated even at late times.
{\it Left:} Evolution of the normalized growing mode $a_\mathcal{R}^{-1} \, {\mathcal{R}}_1$ (red) and decaying mode $a_\mathcal{R}^{-1} \, {\mathcal{R}}_2$ (blue) a few e-foldings before and after horizon exit.
{\it Center:} Evolution in the super-horizon regime. The decaying mode is suppressed by a factor $e^{-3 N_*}$ during inflation and the basis rotation associated with the transition to the RD phase (see text) accounts for an additional suppression of order $\epsilon_*$. During RD, the decaying mode is suppressed by another factor $e^{-N_*}$. {\it Right:} Evolution a few e-foldings before and after horizon re-entry during RD. The decaying mode is rescaled in order to make it visible. In reality it is down by the cumulative suppression factor $\sim \epsilon_* \, e^{-4 N_*} \lesssim 10^{-96}$ (for $N_* \approx 55$). This reflects the standard result that
(ignoring decoherence) the state of the perturbations is extremely squeezed and thus classical in the sense described in the text.
}
\label{fig:SR2RD}
\end{figure*}

While we could now expand the late-Universe operators $\mathcal{R}(\tau)$ and $\mathcal{R}'(\tau)$ in the form of Eq.~(\ref{eq:evol zeta pi}) in terms of the basis of modes $\mathcal{R}_1$ and $\mathcal{R}_2$ in Eqs (\ref{eq:zeta1RD}) and (\ref{eq:zeta2RD}), we still have the freedom of performing a rotation of the form,
\beq
\label{eq:rotmodes}
\begin{pmatrix} \mathcal{R}_1 \\ \mathcal{R}_2 \end{pmatrix} \to \begin{pmatrix} \mathcal{R}_{1'} \\ \mathcal{R}_{2'} \end{pmatrix}  = \begin{pmatrix} \cos \alpha & \sin \alpha \\ -\sin \alpha & \cos \alpha \end{pmatrix} \, \begin{pmatrix} \mathcal{R}_1 \\ \mathcal{R}_2 \end{pmatrix},
\eeq
so that Eq.~(\ref{eq:evol zeta pi}) has the same form in the new basis (see discussion at the end of Section \ref{subsec:sri}).
Since we want the decaying mode to represent the minimal non-commuting component of the operators (see Sections \ref{subsec:squeezing2classical} and \ref{subsec:searching}), we use this freedom to define an infinitesimally rotated basis such that $\mathcal{R}_{2'} \propto \mathcal{R}_{\rm dec}^{\rm RD}$.
In other words, we want this rotation to subtract out the contribution to $\mathcal{R}_2$ proportional to $\mathcal{R}_{\rm grow}^{\rm RD}$. Choosing $\sin\alpha = c/a$, we obtain,
\bea
\label{eq:rot SR}
\mathcal{R}_{1'} &\approx& \mathcal{R}_1 \approx a_\mathcal{R} \, \mathcal{R}_{\rm grow}^{\rm RD}(\tau)  \nonumber \\
\mathcal{R}_{2'} &\approx& \mathcal{R}_2 - \frac{c}{a} \, \mathcal{R}_1 \approx - a_\mathcal{R} \, \frac{\epsilon_*}{6 \sqrt{3}} \, e^{-4 N_*}  \, \mathcal{R}_{\rm dec}^{\rm RD}(\tau)
\eea
(the effect of the rotation on $\mathcal{R}_{1'}$ is suppressed by factors $x_e$),
where the second (approximate) equality on each line gives the leading order\footnote{Note that this result requires expanding the solutions to higher order in $x_e$ than is explicitly written in Eqs (\ref{eq:zeta1RD}) and (\ref{eq:zeta2RD}).} result in $x_e$.
Notice the additional suppression by $\epsilon_*$ above in $\mathcal{R}_{2'}$.

The expansion in terms of late-Universe modes $\mathcal{R}_{1'}$ and $\mathcal{R}_{2'}$ explicitly reads,
\beq
\label{eq:zetahatSR}
\hat{\mathcal{R}}(\tau) =
\sqrt{2} \, \mathcal{R}_{2'}(\tau) \, \hat{x} - \sqrt{2} \, \mathcal{R}_{1'}(\tau) \, \hat{p} =
a_\mathcal{R} \, \left( - \sqrt{2} \, \frac{\epsilon_*}{6 \sqrt{3}} \, e^{-4 N_*} \, {\mathcal{R}}_{\rm dec}^{\rm RD}(\tau) \, \hat{x}
- \sqrt{2} \, {\mathcal{R}}_{\rm grow}^{\rm RD}(\tau) \, \hat{p} \right).
\eeq
(with the Heisenberg-picture state of the perturbations given by the ground state of the annihilation operator, $\hat{a} = \tfrac{1}{\sqrt{2}} \, \hat{x} + \tfrac{\iu}{\sqrt{2}} \, \hat{p}$).
This is the main quantitative result of this Section.
As discussed in Section \ref{subsec:searching}, we consider $\mathcal{R}_{2'}$ (the component proportional to $\hat{p}$ above) to carry the non-commuting ``quantum signature'' of the primordial perturbations.

The evolution of the normalized growing and decaying modes $\mathcal{R}_{1'}$ and $\mathcal{R}_{2'}$ is shown in Figure \ref{fig:SR2RD}.
The left panel shows the evolution during inflation before and slightly after horizon exit. To describe the super-horizon evolution, we switch to a log scale in the middle panel. Finally, the right panel depicts the period slightly before and after horizon entry during RD. At this time, the decaying mode is suppressed by a factor $\sim \epsilon_* \, e^{-4 N_*}$; the right panel therefore shows the decaying mode rescaled by the inverse of this factor.
In the super-horizon regime, we see that, while the growing mode is constant, the decaying mode rapidly decays both during inflation {\it and} during radiation domination. The squeezing of the state continues after inflation is over.
The decaying mode accrues a suppression of $e^{-3 N_*}$ during inflation and another factor
$e^{-N_*}$ during RD. The transition from inflation to radiation domination is responsible for an additional factor of $\epsilon_*$.

We note that, during inflation, the rotated basis $\mathcal{R}_{1'}$, $\mathcal{R}_{2'}$ is indistinguishable from the original basis $\mathcal{R}_1, \mathcal{R}_2$
except near the end of inflation. At this time, $\mathcal{R}_{2'}$ contains a modification relative to $\mathcal{R}_2$ that allows it to evolve directly into the RD decaying mode, cf.~Eq.~(\ref{eq:rot SR}).
This explains the feature seen in $\mathcal{R}_{2'}$ just before the transition from SR to RD.

\vskip 7pt

The main conclusion is that in the standard inflationary scenario,
the decaying mode is suppressed by the factor $\sim \epsilon_* \, e^{-4 N_*}$, which is extremely small (since modes observed in the cosmic microwave background and cosmological large-scale structure typically undergo $N_* \approx 50 - 60$ e-foldings of inflation after exiting the horizon).
Therefore, the quantum signal (in the specific sense of this paper) is extremely small and there is no hope for its detection\footnote{An exception, as long as we are ignoring decoherence, are extremely short modes that exited the horizon very shortly before the end of inflation (and re-entered not long after). These modes have wavelengths a factor $\sim e^{-N_*}$ shorter than the cosmological-scale modes of interest. If a mode with wavelength equal to the Hubble scale today exited the horizon $N \sim 55$ e-foldings before the end of inflation, the short modes undergoing limited squeezing have wavelengths on the order of hundreds of meters}.
The primordial power spectrum is completely dominated by the growing mode and given by the standard expression,
\beq
\label{eq:PPS SR}
\Delta^2_\mathcal{R}(k) \approx \frac{1}{2 \epsilon_* \, m_{\rm pl}^2} \, \left(\frac{H_I}{2 \pi}\right)^2 \, \mathcal{R}_{\rm grow}^{\rm RD \, 2}(\tau).
\eeq

To make the suppression of the decaying mode more quantitative, let us impose that the amplitude of the primordial (i.e.~super-horizon) power spectrum in Eq.~(\ref{eq:PPS SR}) reproduces the observed value.
For simplicity, and since an order-of-magnitude estimate suffices for our purposes,
we still consider our toy model where the post-inflationary Universe is always radiation dominated.
In that case, we have,
\beq
H_I = e^{2 N_*} \, H_{\rm re-entry},
\eeq
where $H_{\rm re-entry}$ is the Hubble parameter at horizon re-entry.
Using the Hubble parameter today for this quantity,
i.e.~$H_0 \approx 70 $km$/$s$/$Mpc,
and inserting the observed value $\Delta^2_\mathcal{R}(k) \approx 2.1 \cdot 10^{-9}$ (at the order-of-magnitude level precision of interest, we do not care that the pivot scale at which the amplitude is measured is not the same as the Hubble scale today), Eq.~(\ref{eq:PPS SR}) then gives,
\beq
\frac{e^{4 N_*}}{\epsilon_*} = 4.1 \cdot 10^{113}.
\eeq
From Eq.~(\ref{eq:zetahatSR}), the suppression factor of the decaying mode amplitude relative to the growing mode amplitude is then,
\beq
\frac{\epsilon_* \, e^{-4 N_*}}{6 \sqrt{3}} \sim 2 \cdot 10^{-115}.
\eeq
The decaying mode is thus suppressed by $\sim 115$ (!) orders of magnitude.

\vskip 21pt

\begin{figure*}[]
\centering
\includegraphics[width=0.5\textwidth]{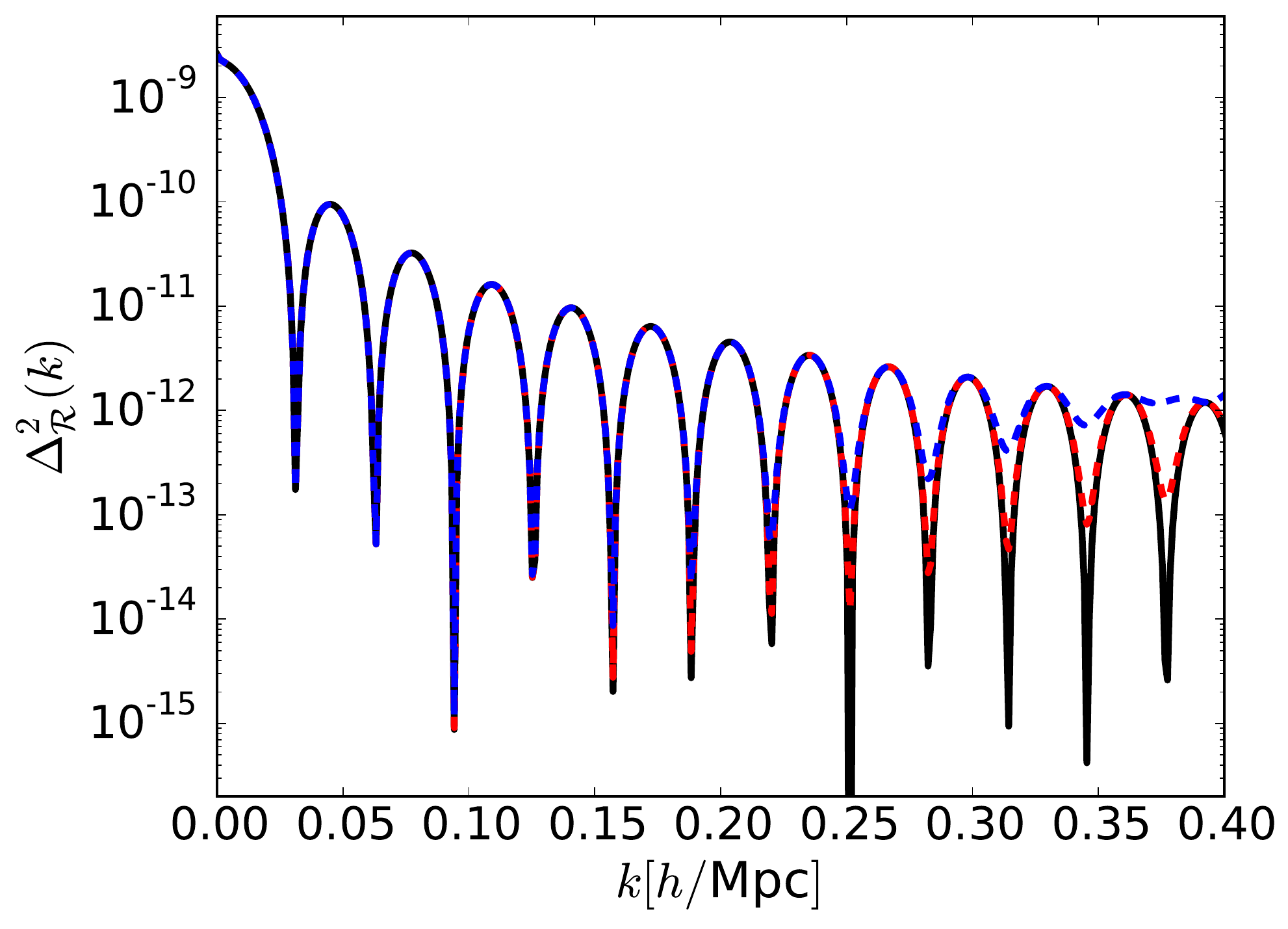} 
\caption{Illustration of effect of the decaying mode on the dimensionless power spectrum of curvature perturbations. We assume a toy model where the late Universe is dominated by a perfect radiation fluid (so that we can use analytic solutions), and the power spectrum is evaluated at a time where the sound horizon $s_{\rm obs} = c_s^{\rm RD} \, \tau_{\rm obs} = 100 \, h^{-1}$Mpc, in order to mimic the acoustic oscillations of the real Universe. We consider a nearly flat primordial spectrum for the growing mode ($n_s = 0.96$) and the decaying mode amplitude is assigned an arbitrary amplitude (cf.~Eqs (\ref{eq:zetahat gen}) and (\ref{eq:Deltadeck})). The spectrum is shown for decaying mode amplitudes $\Delta_{{\rm dec},p} = 0$ (black), $10^{-4}$ (red dashed) and $3 \cdot 10^{-4}$ (blue dashed).
The decaying mode, which encodes the non-commuting component of the operators describing the primordial perturbations, leads to a scale-dependent damping of the acoustic oscillations.
}
\label{fig:PKdecmode}
\end{figure*}

\vskip 7pt

Before asking in the following Sections if there are more exotic inflationary scenarios where the ``quantum contribution'' is {\it not} completely negligible, we illustrate in Figure \ref{fig:PKdecmode} how, {\it if} the decaying mode were indeed non-negligible, it might manifest itself in the power spectrum of $\mathcal{R}$.
Consider a generalization of Eq.~(\ref{eq:zetahatSR}),
\beq
\label{eq:zetahat gen}
\hat{\mathcal{R}}(\tau) = a_\mathcal{R} \, \left( \sqrt{2} \, \Delta_{\rm dec} \, {\mathcal{R}}_{\rm dec}^{\rm RD}(\tau) \, \hat{x}
- \sqrt{2} \, {\mathcal{R}}_{\rm grow}^{\rm RD}(\tau) \, \hat{p} \right),
\eeq
where we will choose the scale-dependence of $a_\mathcal{R}$ such that the growing mode contribution to $\mathcal{R}$ reproduces a nearly scale-invariant, but slightly ``red'' primordial power spectrum,
\beq
\frac{k^3 \, a_\mathcal{R}^2(k)}{2 \pi^2} = 2.1 \cdot 10^{-9} \, \left( \frac{k}{k_p} \right)^{n_s  - 1},
\eeq
with specral index $n_s = 0.96$ and pivot scale $k_p = 0.05 \, h/$Mpc \cite{Aghanim:2018eyx}.
For the decaying mode coefficient, $\Delta_{\rm dec}$, we
assume the same scale-dependence as the actual decaying mode coefficient in Eq.~(\ref{eq:zetahatSR}), $\Delta_{\rm dec} \propto e^{-4 N_*(k)} \propto (k/H_I )^4$, i.e.
\beq
\label{eq:Deltadeck}
\Delta_{\rm dec}(k) \equiv \Delta_{{\rm dec},p} \, \left(\frac{k}{k_p}\right)^4,
\eeq
but $\Delta_{{\rm dec},p}$ now is a free amplitude\footnote{As we discuss in detail in Section \ref{sec:Heis}, if we assume the perturbations are described by a pure quantum state with the operators $\hat{x}$ and $\hat{p}$ as defined in Section \ref{subsec:expvals}, then choosing the decaying mode amplitude to deviate from Eq.~(\ref{eq:zetahatSR}), is actually inconsistent with the canonical commutation relations. Here, we simply illustrate what the effect on the curvature power spectrum would be in a general scenario with non-negligible decaying mode.}, that we allow to be larger than the negligible value in the standard calculation (Eq.~(\ref{eq:zetahatSR})).


We still consider a simplified, radiation-only, post-inflationary Universe, and
treat the radiation as a perfect fluid.
We evaluate the power spectrum at a time $\tau_{\rm obs}$ chosen such that the sound horizon $s_{\rm obs} \equiv c_s^{\rm RD} \, \tau_{\rm obs} = 100 \, h^{-1}\text{Mpc}$, approximately the observed sound horizon at the time of baryon-photon decoupling.
Thus, while the plot lacks many real Universe features, such as those due to baryonic and dark matter, it describes acoustic oscillations similar to those in the actual Universe.
The black curve in Figure \ref{fig:PKdecmode} shows the power spectrum with acoustic oscillations in standard scenario without the decaying mode ($\Delta_{{\rm dec},p} = 0$), while the red and blue curves depict the cases of $\Delta_{{\rm dec},p} = 1 \cdot 10^{-4}$  and $3 \cdot 10^{-4}$.
The decaying mode leads to a damping of the acoustic oscillations because
it reduces coherence in the initial phase of the acoustic oscillations.
This damping has an interesting scale dependence, becoming more pronounced on small scales (large $k$), because $\Delta_{\rm dec}(k) \propto k^4$.


We stress that the decaying mode signature in the power spectrum in itself does not require a quantum mechanical explanation.
The damping of oscillations in the power spectrum due to the decaying mode can easily be described in terms of purely classical stochastic perturbations populating both the growing and decaying mode\footnote{Regardless of the connection to a quantum origin, the amplitude of the decaying mode (distribution of initial phases) can be constrained observationally, as has been done in \cite{amenfin05,Kodwani:2019ynt}.
Note that the description in terms of primordial quantum fluctuations does lead to an interesting specific scale-dependence of the amplitude of the decaying mode relative to the growing mode.}.
As discussed in Section \ref{subsec:searching}, the damping in Figure \ref{fig:PKdecmode} is a quantum signature in the following limited sense.
{\it If} we assume the primordial perturbations are of quantum origin, the damping is a result of the non-commuting (i.e.~decaying mode) component of the quantum operator.
If that damping is large enough to be observable, this non-commuting
component would lead to other signatures in correlation functions beyond the power spectrum that
do deviate from what would be predicted in a classical description.
A final caveat we remind the reader of is that we are using a description where each mode is described as a pure state, independent of other degrees of freedom, while a more realistic description would include the effect of decoherence due to interactions.

\section{Quantum signatures in Ultra-Slow Roll inflation}
\label{sec:USR}

So far we have confirmed the standard result that the decaying mode is completely negligible at late times so that the quantum state of the primordial fluctuations is extremely squeezed and classical.
We now move on and ask if we can construct an inflationary model where the quantum signature is {\it not} entirely negligible in the post-inflationary Universe.
The key reason for the classicalization of the primordial perturbations in the standard scenario was the strong divergence between the growing mode and decaying mode in the super-horizon regime (both during and after inflation).
This motivates us to consider the ultra-slow-roll inflation (USR) scenario, where the roles of the growing and decaying modes are reversed during inflation.
We will in this Section repeat the analysis of the previous Section for the case where the inflationary phase is described by USR and we will compute the resulting quantum signature.
In Section \ref{sec:Heis}, we will then use a more general perspective to explain the results in the USR case and we will use this perspective to draw
more model-independent conclusions about the late-Universe quantum signature.

We again consider the scenario of a phase of inflation, with a (quasi-)de Sitter background, followed by a radiation dominated post-inflationary Universe.
The difference with the standard scenario considered earlier,
is that we now assume that the slow-roll parameter $\epsilon$ evolves with time,
\beq
\epsilon = \epsilon_* \, \left( \frac{a}{a_*}\right)^{\eta},
\eeq
where $a_*$ and $\epsilon_*$ are the scale-factor and slow-roll parameter at the time of horizon exit, $-c_s  k \tau_* = 1$ (we will restrict our attention to the case $c_s = 1$).
The case $\eta = 0$ corresponds to the standard scenario studied above and USR corresponds to a rapidly decreasing slow-roll parameter, $\eta < -3$.
We still assume that, despite the time evolution, during the period of interest, the slow-roll parameter remains small, $\epsilon \ll 1$,
so that the background is consistently approximated by de Sitter.
We set $c_s = 1$ during inflation.

For general $\eta$, the two normalized independent solutions to the equations of motion during inflation, are (cf.~Eq.~(\ref{eq:sols SR norm})),
\bea
\label{eq:sols USR norm}
\mathcal{R}^{\rm USR}_{\text{``grow''}}(\tau) &\equiv& - \sqrt{\frac{\pi}{2}} \, x^{\nu} \, Y_{\nu}(x),  \nonumber \\
\mathcal{R}^{\rm USR}_{\text{``dec''}}(\tau)  &\equiv&   - \sqrt{\frac{\pi}{2}} \, x^{\nu} \, J_{\nu}(x), \quad \text{with} \quad x \equiv - k \, \tau,
\eea
and,
\beq
\nu \equiv \frac{3 + \eta}{2}.
\eeq
The mode $\mathcal{R}^{\rm USR}_{\text{``grow''}}$ approaches a constant on super-horizon scales.
For standard slow-roll inflation, $\mathcal{R}^{\rm USR}_{\text{``dec''}}$ is the decaying mode.
However, for USR inflation, we have $\eta < -3$, and $\mathcal{R}^{\rm USR}_{\text{``dec''}}$ becomes an increasing function of time after horizon exit.
This means $\mathcal{R}^{\rm USR}_{\text{``dec''}}$ becomes the dominant mode instead of $\mathcal{R}^{\rm USR}_{\text{``grow''}}$.
This reversal makes mode evolution in USR qualitatively different from the SR case and is the reason why Maldacena's consistency conditions can be violated in such models.

Assuming the Bunch-Davies vacuum, the mode functions describing evolution of $\hat{\mathcal{R}}(\tau)$ in the Heisenberg picture according to Eq.~(\ref{eq:evol zeta pi}) are,
\bea
\mathcal{R}_1(\tau) &=& a_\mathcal{R} \, \mathcal{R}^{\rm USR}_{\text{``grow''}}(\tau) \nonumber \\
\mathcal{R}_2(\tau) &=& a_\mathcal{R} \, \mathcal{R}^{\rm USR}_{\text{``dec''}}(\tau),
\eea
with
\beq
a_\mathcal{R} \equiv \frac{H_I}{\sqrt{2} \, c_s^{1/2} \, k^{3/2} \, \sqrt{2 \epsilon_*} \, m_{\rm pl}}.
\eeq
Or, in terms of the complex mode function (e.g.~\cite{chenetal13}),
\beq
f(\tau) = \mathcal{R}_2(\tau) + \iu \, \mathcal{R}_1(\tau) =  -\frac{\sqrt{\pi} \, H_I}{2 (c_s \, k)^{3/2}} \, \sqrt{\frac{c_s^2}{2 \epsilon_* \, m_{\rm pl}^2}}\, x^\nu \, H^{(1)}_\nu(x).
\eeq
For slow-roll inflation, the amplitude $a_\mathcal{R}$ coincides with the definition in Eq.~(\ref{eq:norm SR}).
In the case of general $\eta$, $a_\mathcal{R}$ is given by the same expression as in Eq.~(\ref{eq:norm SR}), but with $\epsilon$ specifically evaluated at horizon exit (since $\epsilon$ is now time-dependent).
The basis of solutions of the equation of motion during radiation domination are as given by Eq.~(\ref{eq:modesRD}) in the previous Section.

\begin{figure*}[]
\centering
\includegraphics[width=0.5\textwidth]{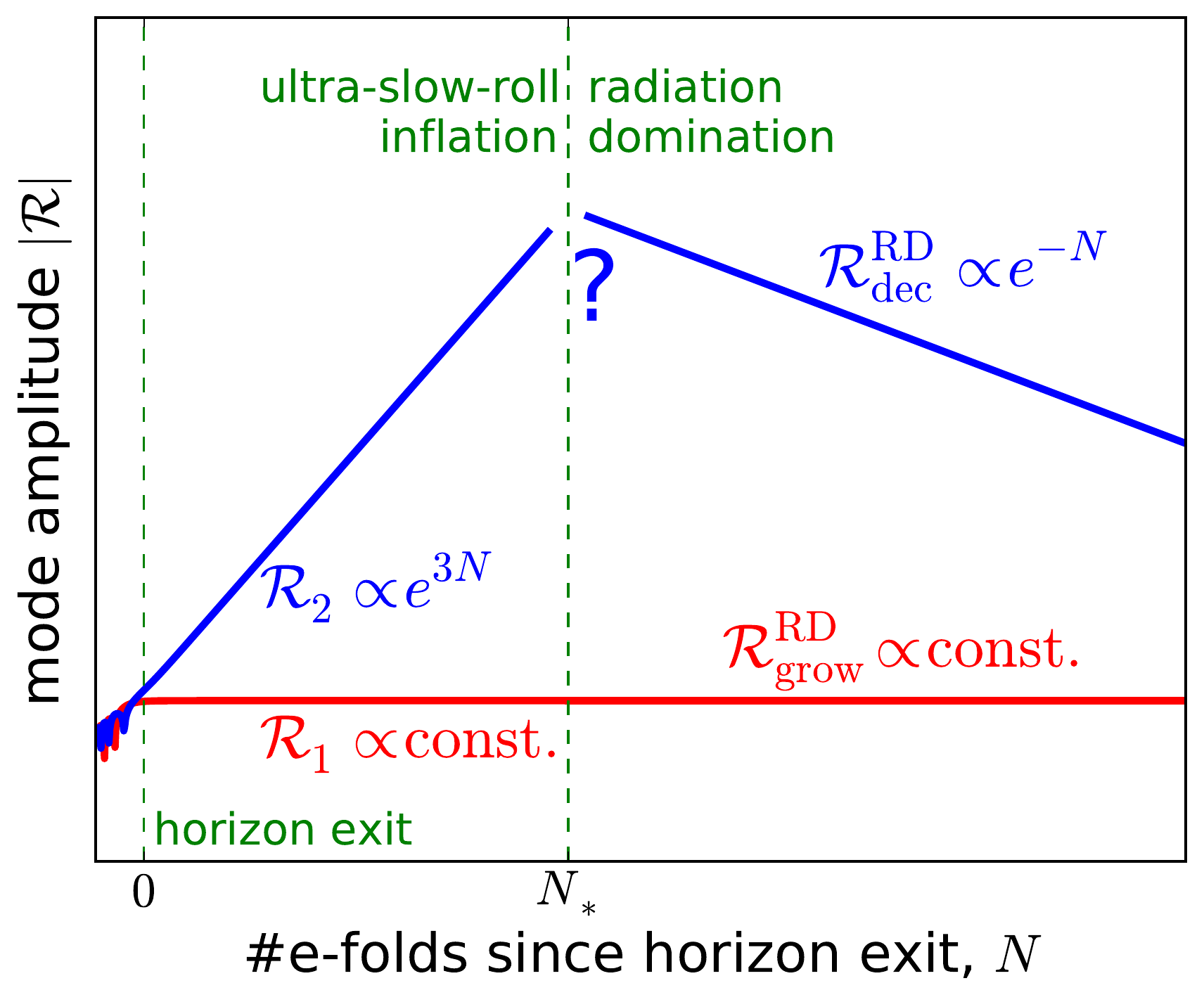} 
\caption{Super-horizon evolution of growing and decaying modes during {\it ultra-slow-roll} (USR) inflation ($N < N_*$)
and radiation domination ($N > N_*$). In the USR scenario, the growing and decaying modes switch roles compared to standard slow-roll inflation: the non-constant mode that would be a decaying mode in SR,
$\mathcal{R}_2 \sim \mathcal{R}^{\rm USR}_{\text{``dec''}}$, grows rapidly in the super-horizon regime, whereas the usual growing mode, $\mathcal{R}_1 \sim \mathcal{R}^{\rm USR}_{\text{``grow''}}(\tau)$, is constant as in SR and evolves into the RD constant mode $\mathcal{R}_{\rm grow}^{\rm RD}$.
The question is then what does the mode $\mathcal{R}^{\rm USR}_{\text{``dec''}}$ evolve into in the RD regime? If it matches onto the RD decaying mode,
as hinted at above (but note the question mark!), one could imagine late-Universe scenarios where the RD decaying mode is not or less suppressed and the state is not fully classical, perhaps leaving an interesting observable signature.
We address whether this is the case in Section \ref{sec:USR} (and Figure \ref{fig:USR2RD}).
}
\label{fig:USR2RDguess}
\end{figure*}

From here on, we will fix $\eta = -6$.
We have above again given the mode functions for general sound speed, but will from here on again restrict discussion to the case of $c_s = 1$.
The resulting scenario of USR inflation followed by radiation domination is intended as a toy model to see if in principle it is possible for the late-Universe decaying mode to {\it not} be heavily suppressed relative to the growing mode. If so, it is worth making the scenario more realistic. One reason the description of inflation in the toy model in its current form is not realistic is that it does not reproduce the correct spectral index $n_s$.
For $\eta = -6$, the super-horizon behavior of the modes to leading order in $- k \tau = |k \, \tau| \ll 1$ is,
\bea
\mathcal{R}^{\rm USR}_{\text{``grow''}}(\tau) &=& \frac{1}{3} + \mathcal{O}((-k \, \tau)^2) \sim \text{const.} \nonumber \\
\mathcal{R}^{\rm USR}_{\text{``dec''}}(\tau) &=& \frac{1}{(-k \, \tau)^3} + \mathcal{O}((-k \, \tau)^{-1})
\sim e^{3 N}.
\eea
The RD modes have the asymptotic behavior,
\bea
{\mathcal{R}}_{\rm grow}^{\rm RD}(\tau) &=& 1 + \mathcal{O}((k \, \tau)^2) \sim \text{const.} \nonumber \\
{\mathcal{R}}_{\rm dec}^{\rm RD}(\tau) &=& \frac{\sqrt{3}}{k \, \tau} + \mathcal{O}(k \, \tau) \sim e^{-N}.
\eea

The question is again what linear combination of ${\mathcal{R}}_{\rm grow}^{\rm RD}$ and ${\mathcal{R}}_{\rm dec}^{\rm RD}$ do the inflationary modes $\mathcal{R}_1 = a_\mathcal{R} \, \mathcal{R}^{\rm USR}_{\text{``grow''}}$ and $\mathcal{R}_2 = a_\mathcal{R} \,\mathcal{R}^{\rm USR}_{\text{``dec''}}$ evolve into?
Just like in the standard slow-roll scenario, we expect (and will soon confirm) that the constant inflationary mode $\mathcal{R}^{\rm USR}_{\text{``grow''}}$ evolves into the constant post-inflationary mode
${\mathcal{R}}_{\rm grow}^{\rm RD}$, i.e.~it simply stays constant on super-horizon scales.
In the SR case, the decaying mode $\mathcal{R}_2 = a_\mathcal{R} \, {\mathcal{R}}_{\rm dec}^{\rm SR}$ (or, really the infinitesimally rotated mode $\mathcal{R}_{2'}$) evolved into the RD decaying mode
${\mathcal{R}}_{\rm dec}^{\rm RD}$, so
one might naively
expect
this to also happen for USR, despite $\mathcal{R}^{\rm USR}_{\text{``dec''}}$ not actually decaying during USR.
If this is the case, then $\mathcal{R}_2$ grows relative to $\mathcal{R}_1$ during USR inflation, but would decay afterwards.
In that hypothetical scenario, by the time the perturbations are observed (say, around the time of horizon re-entry),
the decaying mode would dominate over the growing mode.
We illustrate this in Figure \ref{fig:USR2RDguess}.
If this is the case, one could then easily imagine tweaking this model to create a late-Universe state where the two modes are of the same order at the time of observation.
Therefore, in this scenario, one could perhaps generate perturbations that have a non-negligible remaining quantum signature.
This is the motivation for studying the USR toy model in this Section.

\vskip 7pt

We now look quantitatively what happens by actually evolving the inflationary modes ${\mathcal{R}}_1$, ${\mathcal{R}}_2$ through the USR-to-RD transition, analogously to the analysis in the previous Section.
The matching conditions are,
\bea
\label{eq:matchingUSR}
\mathcal{R}_+ &=& \mathcal{R}_- \quad \text{(continuous)} \nonumber \\
\mathcal{R}'_+ &=& \frac{\left( 2 \epsilon/c_s^2 \right)_-}{\left( 2 \epsilon/c_s^2 \right)_+} \, \mathcal{R}_-'
= \frac{\epsilon_e}{6} \, \mathcal{R}_-' = \frac{\epsilon_*}{6} \, e^{-6 N_*} \, \mathcal{R}_-',
\eea
The discontinuity in $\mathcal{R}'$ is again proportional to the slow-roll parameter $\epsilon$ at the time of the transition (i.e.~at the end of inflation). In USR, because $\epsilon$ decays as $\epsilon \propto a^{-6}$ through the many e-foldings of inflation,
$\epsilon_e$ is strongly suppressed relative to the value at horizon exit, $\epsilon_*$, which is itself assumed to be small.

We now again write the behavior of $\mathcal{R}_1$ and $\mathcal{R}_2$ during RD as,
\bea
\mathcal{R}_1(\tau) &=& a \, \mathcal{R}_{\rm grow}^{\rm RD}(\tau) + b \, \mathcal{R}_{\rm dec}^{\rm RD}(\tau) \quad  \text{\bf (during RD)} \nonumber \\
\mathcal{R}_2(\tau) &=& c \, \mathcal{R}_{\rm grow}^{\rm RD}(\tau) + d \, \mathcal{R}_{\rm dec}^{\rm RD}(\tau) \quad  \text{\bf (during RD)}.
\eea
Following the same procedure as in Section \ref{sec:SR2RD}, but now with the matrix describing $\mathcal{R}_1$ and $\mathcal{R}_2$ just after the transition given by,
\beq
{\bf E}_{1,2} \equiv \begin{pmatrix} \, \mathcal{R}_1(\tau_e)  & \, \mathcal{R}_2(\tau_e) \\ \mathcal{R}_1'(\tau_e)  & \, \, \mathcal{R}_2'(\tau_e) \end{pmatrix}
= a_\mathcal{R} \, \begin{pmatrix} \, \mathcal{R}_{\text{``grow''}}^{\rm USR}(-\tau_e)  & \, \mathcal{R}_{\text{``dec''}}^{\rm USR}(-\tau_e) \\ \epsilon_e/6 \, (\mathcal{R}_{\text{``grow''}}^{\rm USR})'(-\tau_e)  & \, \, \epsilon_e/6 \, (\mathcal{R}_{\text{``dec''}}^{\rm USR})'(-\tau_e) \end{pmatrix},
\eeq
we find
to leading order in $x_e = k \, \tau_e = e^{-N_*} \ll 1$,
\bea
\label{eq:USR2SRmodes}
a_\mathcal{R}^{-1} \, {\mathcal{R}}_1(\tau)
&=& \left(\frac{1}{3} + \mathcal{O}(x_e^2) \right) \, {\mathcal{R}}_{\rm grow}^{\rm RD}(\tau)
+ \left( - \frac{1}{27 \sqrt{3}} \, x_e^3 + \mathcal{O}(x_e^5) \right) \, {\mathcal{R}}_{\rm dec}^{\rm RD}(\tau) \quad \text{\bf (during RD)} \nonumber \\
a_\mathcal{R}^{-1} \, {\mathcal{R}}_2(\tau)
&=& \left( x_e^{-3} + \mathcal{O}(x_e^{-1})\right) \, {\mathcal{R}}_{\rm grow}^{\rm RD}(\tau)
+ \left( -\frac{1}{9 \sqrt{3}} + \mathcal{O}(x_e^2) \right) \, {\mathcal{R}}_{\rm dec}^{\rm RD}(\tau) \quad \text{\bf (during RD)}.
\eea
As expected, we see that the constant mode ${\mathcal{R}}_1$ evolves into the constant, RD growing mode, modulo tiny corrections of order $x_e^2 = e^{-2 N_*}$.
A major difference with the SR scenario is that after the end of inflation, the mode ${\mathcal{R}}_2$ is now the dominant mode, with amplitude a factor $\sim x_e^{-3} = e^{3 N_*}$ above ${\mathcal{R}}_1$.
It evolves predominantly into the RD growing mode. In the SR scenario, it was also the case that both modes mostly map into ${\mathcal{R}}_{\rm grow}^{\rm RD}$, but in the present case, the RD decaying mode contribution to ${\mathcal{R}}_2$ at $\tau_e$ is suppressed by $x_e^2 = e^{-2 N_*}$ instead of the more modest factor of $\epsilon_*$ in the SR case.

\begin{figure*}[]
\centering
\includegraphics[width=0.32\textwidth,height=0.25\textwidth]{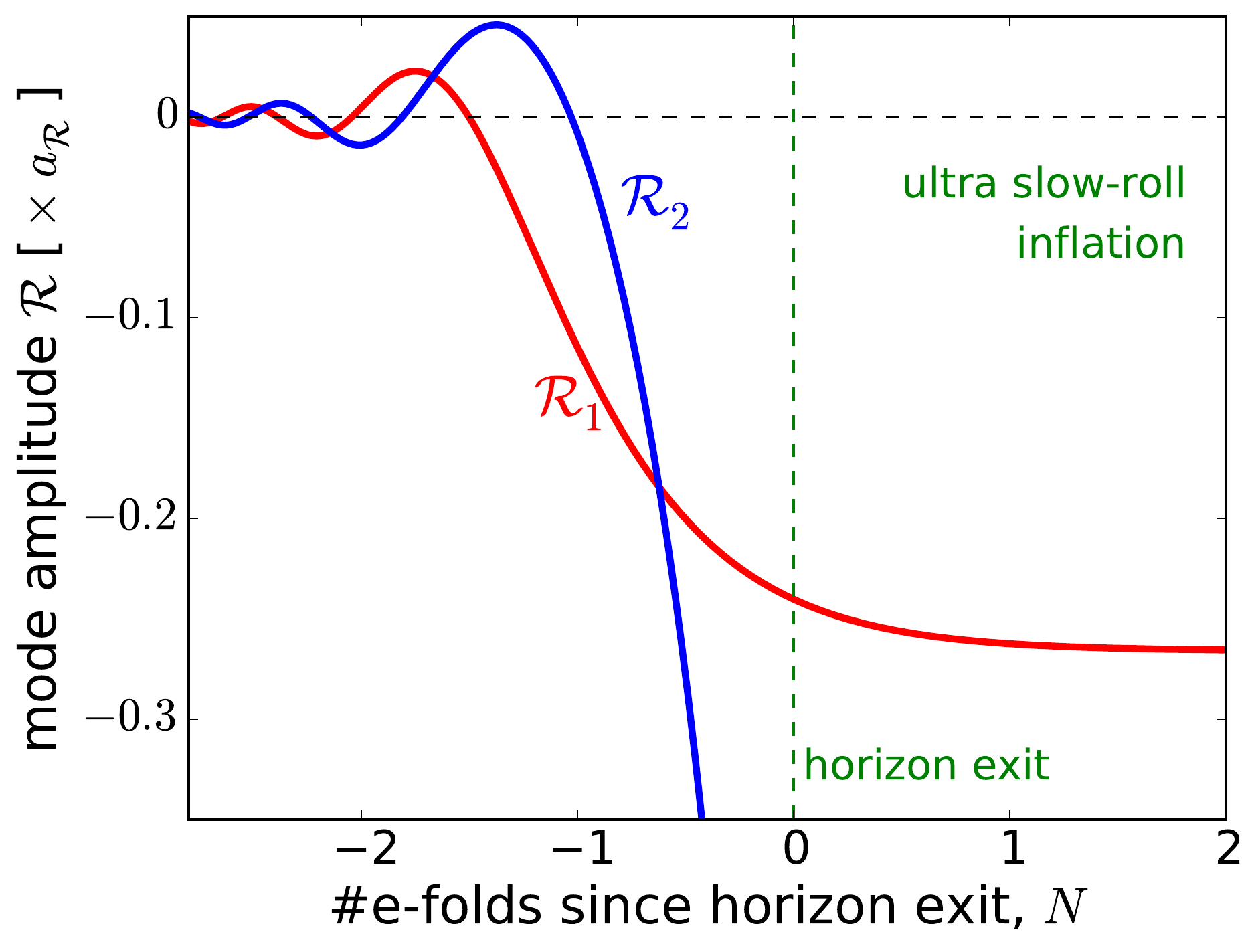} 
\includegraphics[width=0.32\textwidth,height=0.25\textwidth]{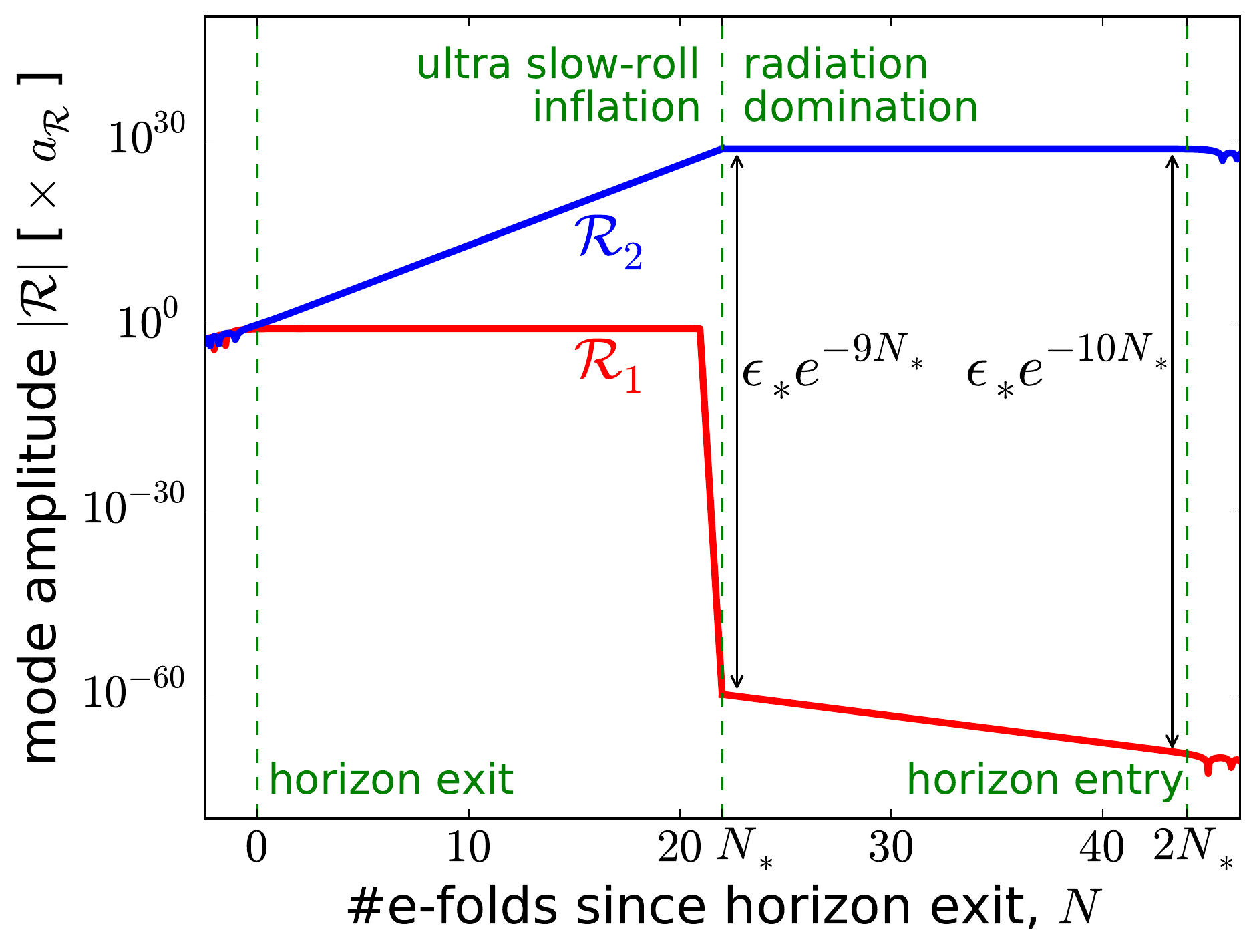} 
\includegraphics[width=0.32\textwidth,height=0.25\textwidth]{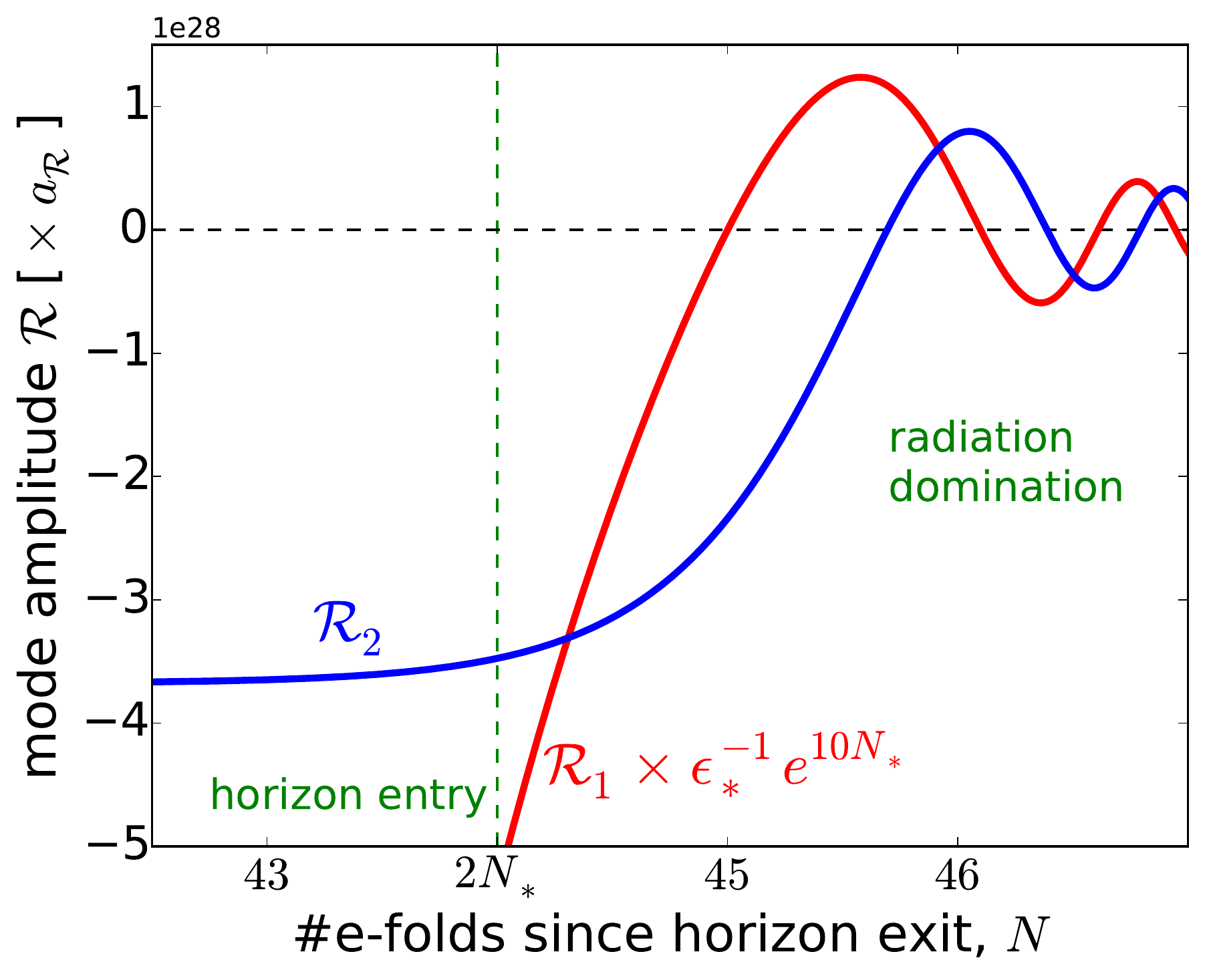} 
\caption{As Figure \ref{fig:SR2RD}, but with the inflationary phase governed by the {\it ultra-slow-roll} scenario.
Explicit calculation of the evolution of the inflationary mode ${\mathcal{R}}_2$ into the RD regime shows shows that it evolves mostly into the RD growing mode (unlike the hypothetical scenario depicted with the question mark in Figure \ref{fig:USR2RDguess}). Expanding in terms of growing and decaying modes with respect to late-Universe behavior, we find that at the start of the RD phase, the RD decaying mode is suppressed by a factor $\epsilon_* \, e^{-9 N_*}$,
of which $e^{-3 N_*}$ is due to the growth of ${\mathcal{R}}_2$ during USR inflation, and the remaining
$\epsilon_* \, e^{-6 N_*}$ due to the basis transformation associated with the transition at the end of inflation, leading the mode ${\mathcal{R}}_{1'}$ to evolve into the RD decaying mode.
As in the standard scenario, super-horizon evolution after inflation contributes another $e^{- N_*}$ suppression.
Thus, even in the USR scenario, the resulting state of the perturbations is extremely squeezed and quantum effects of the type discussed in the text are hopelessly suppressed.
We explain in Section \ref{sec:Heis} that, in single-field inflation, this will in fact be the case no matter what exotic scenario we consider to modify the behavior of the growing and decaying modes.
}
\label{fig:USR2RD}
\end{figure*}

We now again apply a rotation in the basis of modes so that $\hat{\mathcal{R}}(\tau)$ is expanded in terms of late-Universe (RD) growing and decaying modes (cf.~Eq.~(\ref{eq:rotmodes}) and surrounding discussion), maximizing the hierarchy between the two modes.
It is then that the subdominant/decaying mode can be interpreted as the minimal non-commuting component, or {\it quantum} component, as dicussed in Section \ref{subsec:searching}.
Concretely, we here apply a rotation that subtracts the ${\mathcal{R}}_{\rm grow}^{\rm RD}$ contribution from the subdominant mode ${\mathcal{R}}_1$, i.e.~$\sin \alpha = -a/c$, giving\footnote{This calculation requires expanding the mode coefficients to higher order in $x_e$ than what is explicitly written in Eq.~(\ref{eq:USR2SRmodes}).},
\bea
\mathcal{R}_{1'}(\tau) &\approx& {\mathcal{R}}_1(\tau) - \frac{a}{c} \, {\mathcal{R}}_2(\tau)
\approx a_\mathcal{R} \, \frac{\epsilon_*}{6 \sqrt{3}} \, e^{-7 N_*} \, {\mathcal{R}}_{\rm dec}^{\rm RD}(\tau) \nonumber \\
{\mathcal{R}}_{2'}(\tau) &\approx& {\mathcal{R}}_{2}(\tau) \approx a_\mathcal{R} \, e^{3 N_*} \, {\mathcal{R}}_{\rm grow}^{\rm RD}(\tau),
\eea
where the effect of the infinitesimal rotation on ${\mathcal{R}}_{2'}$ is negligible.
The mode $\mathcal{R}_{2'}(\tau)$ is thus the dominant, growing mode in the late Universe and $\mathcal{R}_{1'}(\tau)$ is the decaying mode.
This is a reversal with respect to the slow-roll scenario.
Note, however, that we could easily apply an additional $90$ degree rotation ($\mathcal{R}_{1'} \to \mathcal{R}_{2'}$, $\mathcal{R}_{2'} \to -\mathcal{R}_{1'}$) if we wanted to match the convention that $\mathcal{R}_{1'}$ is the late-Universe growing mode. We do not do this here.
The curvature perturbation operator in terms of late-Universe growing and decaying modes thus reads,
\bea
\label{eq:zetahatUSR}
\hat{\mathcal{R}}(\tau) =
\sqrt{2} \, \mathcal{R}_{2'}(\tau) \, \hat{x} - \sqrt{2} \, \mathcal{R}_{1'}(\tau) \, \hat{p}
&=& \frac{H_I}{\sqrt{2} \,  k^{3/2} \, \sqrt{2 \epsilon_*} \, m_{\rm pl}} \,
\left( e^{3 N_*} \, {\mathcal{R}}_{\rm grow}^{\rm RD}(\tau) \, \hat{x}
- \frac{\epsilon_*}{6 \sqrt{3}} \, e^{-7 N_*} \, {\mathcal{R}}_{\rm dec}^{\rm RD}(\tau) \, \hat{p} \right) \nonumber \\
&=& \frac{H_I}{\sqrt{2} \, k^{3/2} \, \sqrt{2 \epsilon_e} \, m_{\rm pl}} \, \left( {\mathcal{R}}_{\rm grow}^{\rm RD}(\tau) \, \hat{x}
- \frac{\epsilon_e}{6 \sqrt{3}} \, e^{-4 N_*} \, {\mathcal{R}}_{\rm dec}^{\rm RD}(\tau) \, \hat{p} \right).
\eea
This is the main quantitative result of this Section.

We show the evolution of the growing mode ${\mathcal{R}}_{2'}$ and the decaying mode ${\mathcal{R}}_{1'}$, the latter encoding the leftover ``quantum'' signal, in Figure \ref{fig:USR2RD}.
The left panel shows the period around horizon exit during inflation. The mode ${\mathcal{R}}_{2'}$ rapidly grows after horizon exit $\propto e^{3 N}$, and the mode $\mathcal{R}_{1'}$ reaches a constant.
The middle panel shows the ensuing super-horizon behavior and the transition into the (super-horizon) RD regime.
Unlike the previously expressed naive expectation, the non-constant mode ${\mathcal{R}}_{2'}$ does not evolve into the RD decaying mode, but transitions into the post-inflationary growing mode, which is constant\footnote{It is worth noting here that the USR super-horizon evolution of the dominant mode is very different for
the curvature perturbation on constant energy density hypersurfaces, $\zeta$, than for $\mathcal{R}$. While the latter grows as $\mathcal{R}_{\rm ``dec''}^{\rm USR} \propto e^{3 N}$, the former only grows as $\zeta_{\rm ``dec''}^{\rm USR} \propto e^{N}$ (see e.g.~\cite{romanomooijsasaki16}). This means that at the end of the USR phase, the amplitude of $\zeta$ is highly suppressed relative to $\mathcal{R}$. Interestingly, this mismatch is compensated for by the transition. Our matching conditions of continuous $\mathcal{R}$ imply a discontinuity in $\zeta$. This step in $\zeta$ assures that, to leading order in gradients, $\zeta$ and $\mathcal{R}$ coincide in the RD epoch. This is as it should be because during RD, the (dominant mode of the) perturbation occupies the constant mode.}.
While the constant mode of the inflationary era, ${\mathcal{R}}_{1}$, also predominantly evolves into the RD growing mode, the mode in the rotated basis that evolves into the post-inflationary decaying mode, i.e.~${\mathcal{R}}_{1'}$, is a linear combinaton of ${\mathcal{R}}_{1}$ and ${\mathcal{R}}_{2}$. This linear combination is equal to ${\mathcal{R}}_{1}$ until close to the end of inflation, but just before the transition gets suppressed strongly to match onto the RD decaying mode with the appropriate amplitude.
Thus, {\it during} inflation, the decaying mode originally gets suppressed relative to the growing mode by a factor $e^{-3 N_*}$, but
the transition leads to another factor $\epsilon_* \, e^{-6 N_*} = \epsilon_e$.
Finally, during the RD phase, the suppression is exacerbated by the standard additional factor of $e^{-N_*}$, leading to a final suppression at horizon entry of, $\epsilon_* \, e^{-10 N_*}$.
The right panel shows the evolution slightly before and after horizon entry. Since the decaying mode is heavily suppressed, we have multiplied it by a factor $\epsilon_*^{-1} \, e^{10 N_*}$ to compensate.


\vskip 7pt

In the USR scenario, around the time of horizon re-entry, the decaying mode amplitude is suppressed by a factor $\sim \epsilon_*^{-1} \, e^{- 10 N_*}$ relative to the growing mode.
This appears to be an even stronger suppression than the factor $\sim \epsilon_*^{-1} \, e^{- 4 N_*}$ we found in the SR scenario.
However, we should
compare the two scenarios for {\it fixed}
primordial amplitude\footnote{We remind the reader that the ultra-slow-roll scenario is merely a toy model and that, even if we tune parameters to reproduce the observed primordial scalar amplitude $A_s$, the model in its current simple form is inconsistent with other observations such as the scalar index $n_s$ of the primordial power spectrum.},
to the observed value $A_s = 2.1 \cdot 10^{-9}$ \cite{Aghanim:2018eyx}.
For USR, the super-horizon
amplitude of the primordial power spectrum is (see Eq.~(\ref{eq:zetahatUSR})),
\beq
\Delta_\mathcal{R}^2(k) \approx
\frac{1}{2 \epsilon_e \, m_{\rm pl}^2} \, \left(\frac{H_I}{2 \pi}\right)^2
= \frac{e^{6 N_*}}{2 \epsilon_* \, m_{\rm pl}^2} \, \left(\frac{H_I}{2 \pi}\right)^2.
\eeq
Imposing the observed primordial amplitude according to the discussion below Eq.~(\ref{eq:PPS SR}), for USR we then have in our toy model,
\beq
\frac{e^{10 N_*}}{\epsilon_*} = 4.1 \cdot 10^{113}.
\eeq
For SR inflation, the primordial amplitude instead imposed $e^{4 N_*}/\epsilon_* = 4.1 \cdot 10^{113}$, so that the decaying mode suppression factors in the two scenarios are in fact equal.
Concretely, the USR suppression factor is (cf.~Eq.~(\ref{eq:zetahatUSR})),
\beq
\frac{\epsilon_* \, e^{-10 N_*}}{6 \sqrt{3}} \sim 2 \cdot 10^{-115}.
\eeq
Therefore, by requiring that the primordial power spectrum amplitude is the same,
we have found that the suppression of the decaying mode relative to the growing mode is the same $\sim 115$ orders of magnitude in both scenarios.

Finally, note from the above discussion that, for fixed value of $\epsilon_*$, the USR scenario requires a smaller number of inflationary e-foldings to reproduce the same primordial amplitude, because,
\beq
\left( \frac{e^{4 N_*}}{\epsilon_*} \right)_{\rm SR} =  \left( \frac{e^{10 N_*}}{\epsilon_*} \right)_{\rm USR}.
\eeq
This is (approximately) reflected on the horizontal axes of Figure \ref{fig:USR2RD} by the short duration of inflation relative to the SR plots.

In summary,
while we might have hoped for the hierarchy between growing and decaying mode to be less dramatic in the USR toy model than in the standard SR scenario, perhaps leaving room for a non-negligible decaying mode in a more realistic version of the scenario,
we find that,
if we choose the USR parameters such that they reproduce the same
(growing mode) primordial power spectrum amplitude as in the SR case, the suppression of the decaying mode turns out to be identical to the suppression seen in the SR scenario.
We will show in Section \ref{sec:Heis} that this is not a coincidence, but instead reflects a very general property of the primordial perturbations
related to Heisenberg's uncertainty principle.

\section{General suppression of quantum signatures and Heisenberg's uncertainty principle}
\label{sec:Heis}

We have shown that even in the unorthodox USR inflation toy model of Section \ref{sec:USR}, the late-Universe state of the primordial perturbations is highly squeezed, and the ``non-classical'' contribution extremely suppressed.
We will now show that, under a simple set of assumptions, this type of cosmic quantum censorship holds generally.
The argument relies on two main assumptions. First, we assume that the late-Universe curvature perturbations are described by a pure, Gaussian state. This assumption is satisfied if the initial state is Gaussian, all subsequent evolution is linear, and there is no mixing with additional degrees of freedom such as isocurvature perturbations. In reality, the perturbations of course undergo interactions with various environments, leading to decoherence, but it is still interesting to consider the idealized scenario where this is neglected.
The second assumption we will use is the observational fact that the perturbations are dominated by the late-Universe growing mode, with amplitude given by the measured value of the primordial scalar amplitude $A_s$.

\subsection{Derivation in terms of mode functions}
\label{subsec:Wronski}

Under the Gaussian assumption, in the Heisenberg picture, we write as before (see Eq.~(\ref{eq:evol zeta pi}) and surrounding discussion),
\beq
\hat{\mathcal{R}}(\tau) = \sqrt{2} \, \mathcal{R}_2(\tau) \, \hat{x} - \sqrt{2} \, \mathcal{R}_1(\tau) \, \hat{p},
\eeq
where $\hat{x}$ and $\hat{p}$ are time-independent operators acting on the harmonic oscillator vacuum.
Since we can always apply a rotation to the operators $\hat{x}$ and $\hat{p}$, we without loss of generality assume that $\mathcal{R}_1(\tau)$ is the dominant mode, which because of our second assumption above is mainly on the late-Universe growing mode, ${\mathcal{R}}_{\rm grow}^{\rm RD}(\tau)$, and that $\mathcal{R}_2(\tau)$ is proportional to the late-Universe decaying mode, ${\mathcal{R}}_{\rm dec}^{\rm RD}(\tau)$.
Thus, we explicitly write
the dimensionless mode functions,
\beq
\label{eq:ansatz modes}
\sqrt{\frac{k^3}{2 \pi^2}}\, \mathcal{R}_1(\tau) = a_1  \, {\mathcal{R}}_{\rm grow}^{\rm RD}(\tau) + b_1 \, {\mathcal{R}}_{\rm dec}^{\rm RD}(\tau),
\quad \sqrt{\frac{k^3}{2 \pi^2}}\,  \mathcal{R}_2(\tau) = a_2 \, {\mathcal{R}}_{\rm dec}^{\rm RD}(\tau),
\eeq
where $a_1, a_2$ and $b_2$ are dimensionless coefficients,
and,
\beq
a_1^2 \approx A_s \approx 2.1 \cdot 10^{-9}.
\eeq
The mode $\mathcal{R}_1(\tau)$ describes the ``classical'', commuting component of the perturbations $\hat{\mathcal{R}}$ and $\hat{\Pi}$, and $\mathcal{R}_2(\tau)$ the ``quantum'', non-commuting contribution.
While observations thus already tell us the quantum contribution is subdominant, the question of interest in this paper has been and is whether it might be large enough in certain scenarios to be in principle detectable.
We will in this Section keep explicit factors of $\hbar$ (while still working in units with $c = 1$).

One way of proving that the decaying mode will in general be extremely suppressed is to use the Wronskian of the equations of motion.
For the action (\ref{eq:action}), the conserved Wronskian is,
\beq
\label{eq:def Wronskian}
W \equiv z^2(\tau) \, \left( \mathcal{R}_1(\tau) \, \mathcal{R}_2'(\tau) - \mathcal{R}_1'(\tau) \,  \mathcal{R}_2(\tau) \right) = \frac{\hbar}{2},
\eeq
where during RD,
\beq
z(\tau) =  \sqrt{\frac{2 \epsilon_{\rm RD}}{c_{s,{\rm RD}}^2}} \, \hbar^{-1/2} \, m_{\rm pl} \, a(\tau) = 2 \sqrt{3} \, \hbar^{-1/2} \, m_{\rm pl} \, a(\tau).
\eeq
Substituting Eq.~(\ref{eq:ansatz modes}) into the Wronskian and using $\tau = 1/(a H(a))$ gives,
\beq
a_1 \, a_2 = \frac{1}{48 \sqrt{3} \, \pi^2} \, \frac{\left( k/a \right)^4}{\left( m_{\rm pl}/\hbar\right)^2 \, H^2(a)}.
\eeq
Thus, the decaying mode is suppressed by,
\beq
\label{eq:suppression mode amplitudes}
\frac{a_2}{a_1} = \frac{1}{48 \sqrt{3} \, \pi^2} \, \left(\frac{k/a}{H(a)}\right)^4 \, \left(\frac{H(a)}{m_{\rm pl}/\hbar}\right)^2 \, A_s^{-1}.
\eeq
Considering modes with physical wave number $k/a$ of order (or within a few orders of magnitude of) the Hubble scale, we see that the origin of the large suppression of the decaying mode is the hierarchy between
the present Hubble scale and the Planck scale. While the dimensionless growing mode amplitude is small, $A_s \approx 2.1 \cdot 10^{-9}$, this does not come close to compensating for the factor,
\beq
\label{eq:suppression}
\left(\frac{H(a)}{m_{\rm pl}/\hbar}\right)^2 \sim \left(\frac{H_0}{m_{\rm pl}/\hbar}\right)^2 \approx 4 \cdot 10^{-121}.
\eeq
Indeed, inserting $k/a = H(a)$ and $H(a) = H_0$ into Eq.~(\ref{eq:suppression mode amplitudes}) reproduces the suppression factor,
\beq
\frac{a_2}{a_1} \approx 2 \cdot 10^{-115},
\eeq
found in the explicit calculations of the SR and USR scenarios.


\subsection{Derivation in terms of Wigner function}
\label{subsec:deriv wig}


The late-time classicality of the perturbations can be understood in an alternative way that allows an analogy to the existence of very classical, coherent states for a particle with macroscopic mass $M$.
Consider a particle of which the action has the kinetic term,
\beq
S_K = \frac{1}{2} \, \int dt \, M \, \dot{x}^2(t),
\eeq
where $x(t)$ is its position as a function of time and the dot denotes a time derivative.
Its conjugate momentum is then,
\beq
p = M \, \dot{x}.
\eeq
Heisenberg's uncertainty principle states that,
\beq
\sigma(x) \, \sigma(p) \geq \frac{\hbar}{2},
\eeq
which is minimally satisfied for a coherent state.
Translating this into the uncertainties in $x$ and $\dot{x}$ for such a state, gives,
\beq
\label{unc part x xdot}
\sigma(x) \, \sigma(\dot{x}) = \frac{\hbar}{2 M}.
\eeq
Since $\hbar \approx 10^{-34} kg \, m^2/s$,
if the mass $M$ is macroscopic, say of order kilograms,
then the right-hand side of the above equation is tiny in macroscopic units of seconds and meters.
This means that coherent states exist for the particle such that the uncertainty in both the position and the velocity are extremely small compared to any macroscopic scale. In a sense, this is just a convoluted way of stating that $\hbar$ is a very small number in macroscopic units\footnote{We stress that the discussion of the masssive particle does not show that states with macroscopically large quantum uncertainty do not exist at all for  macroscopic objects. We are simply discussing the position and momentum uncertainties in {\it coherent} states. An explanation of why general macroscopic quantum superpositions (e.g.~superpositions of two coherent states with macroscopically different mean positions) are very difficult to obtain, necessarily involves decoherence.}.
The reason it was useful to spell this out, however, is that we can now make the analogy with the primordial cosmic perturbations.

The generalization to general Gaussian states (and in particular to squeezed states of primordial perturbations)
is
that the {\it area} inside the Wigner ellipse is set by $\hbar/2$. In the phase space given by a variable, e.g.~$\mathcal{R}$, and its conjugate momentum, the area enclosed (in a $\chi^2 = 1$ contour) is\footnote{This area is given by $\text{Area}(\mathcal{R}, \Pi) = \pi \sqrt{\text{Det}{\bf C}}$, where ${\bf C}$ is the covariance matrix describing the Wigner funcion, Eq.~(\ref{eq:covWig}). The quantity $\sqrt{\text{Det}{\bf C}}$ is exactly equal to the Wronskian, Eq.~(\ref{eq:def Wronskian}),
and conservation of the Wronskian is thus equivalent to conservation of the area of the Wigner ellipse.},
\beq
\text{Area}(\mathcal{R}, \Pi) = \pi \, \frac{\hbar}{2}.
\eeq
As illustrated in Figure \ref{fig:wigner illus} and discussed in Sections \ref{subsec:squeezing2classical} and \ref{subsec:searching}, this area can be seen as the product of the extent of the stretched ``classical'' (growing mode) direction in phase space and the squeezed ``quantum'' (decaying mode) direction.
To determine this area directly in units relevant for cosmological observations, consider the phase space defined by the dimensionless curvature perturbation $\sqrt{k^3/2\pi^2} \, \mathcal{R}$ (with variance equal to the dimensionless power spectrum $\Delta_\mathcal{R}^2(k)$)
and its rate of change per Hubble time, i.e.~we use $N = \ln a$ as the time coordinate.
Then, following the example of the massive particle, we write the kinetic part of the action Eq.~(\ref{eq:action}) as,
\beq
\label{eq:action SK}
S_K = \frac{1}{2} \, \int d\ln a \, M_\mathcal{R}(\tau) \, \left| \sqrt{\frac{k^3}{2 \pi^2}} \, \frac{d \mathcal{R}}{d \ln a} \right|^2,
\eeq
with an ``effective mass'' (note, however, that $M_\mathcal{R}$ does not have units of mass),
\beq
\label{eq:eff mass zeta}
M_\mathcal{R}(\tau) \equiv \frac{2 \pi^2 \, z^2(\tau)}{k^3 \, \tau} = \frac{24 \pi^2}{\hbar} \, \left(\frac{m_{\rm pl}/\hbar}{H(a)}\right)^2 \, \left(\frac{H(a)}{k/a}\right)^3,
\eeq
where the second equality holds
during radiation domination. A key realization is that this effective mass is extremely large (in units $\hbar$) in the late Universe due to the factor $(m_{\rm pl}/H(a))^2 \sim 10^{-121}$.
Analogously to Eq.~(\ref{unc part x xdot}), we now obtain the generalized uncertainty relation for a squeezed, Gaussian state,
\beq
\label{eq:area dimless}
\left[ \Delta^2_\mathcal{R}(k) \right]^{1/2} \, \left[\Delta^2_{\delta (d\mathcal{R}/d\ln a)_{\rm qu}}(k)\right]^{1/2} \approx \pi^{-1} \, \text{Area}\left(\sqrt{\frac{k^3}{2 \pi^2}} \, \mathcal{R}, \sqrt{\frac{k^3}{2 \pi^2}} \, \frac{d \mathcal{R}}{d \ln a} \right) = \frac{\hbar}{2 M_\mathcal{R}}.
\eeq
where the quantum component $\delta (d\mathcal{R}/d\ln a)_{\rm qu}$ quantifies the deviation of $d\mathcal{R}/d\ln a$ from the ``classical'' growing mode solution and scales as $\delta (d\mathcal{R}/d\ln a)_{\rm qu} \propto \mathcal{R}_2'(\tau)$.

Since $\hbar/M_\mathcal{R}(\tau)$ is extremely small, it follows from Eq.~(\ref{eq:area dimless}) that the product of the extents of the classical and quantum directions in phase space is infinitesimal. Since the classical direction is measured, $ \Delta^2_\mathcal{R}(k)  \approx A_s$, it again follows that the quantum direction (above quantified by $\Delta^2_{\delta (d\mathcal{R}/d\ln a)_{\rm qu}}(k)$) is unmeasurably small.
This is exactly the same suppression of the decaying mode that was found using conservation of the Wronskian in Eq.~(\ref{eq:suppression mode amplitudes}), as can be explicitly checked by expressing the area in the Wigner ellipse in terms of the mode functions $\mathcal{R}_1$, $\mathcal{R}_2$.

\subsection{Discussion}

The suppression of the ``quantum'' decaying mode found above, Eqs (\ref{eq:suppression mode amplitudes}) or (\ref{eq:area dimless}), exactly reproduces the suppression found for the specific slow-roll and ultra slow-roll toy models in Sections \ref{sec:USR} and \ref{sec:SR2RD}.
We have shown that this is a quite general result that follows directly from the (kinetic part of the) action for the cosmic curvature perturbations in the late Universe.
The characteristic energy scale appearing in this action is the Planck scale, $m_{\rm pl}$, while the length and time scales relevant for cosmic perturbations in the present Universe are of order the Hubble scale\footnote{This is in practice true up to a few orders of magnitude. We could of course consider observations at the time of recombination, where the Hubble scale is larger than today, and in general we also observe modes smaller than the Hubble scale. However, such refinements cause only a small modification to the number of orders magnitude by which the scales of observational relevance are different from the Planck scale.}, $H_0$.
This means that if the perturbations and time coordinate are expressed in the units natural to cosmological observations, the action is analogous to that of a macroscopic particle with large mass, $M_\mathcal{R} \sim (m_{\rm pl}/H_0)^2$. By a version of Heisenberg's uncertainty principle applied to general Gaussian states, this large mass implies an extremely small product of the stretched ``classical'' and squeezed ``quantum'' directions in phase space.

Interestingly, the ratio $(H_0/m_{\rm pl})^2$ that suppresses the decaying mode is the same ratio appearing in the cosmological constant problem, $\rho_\Lambda/m_{\rm pl}^4 \sim \rho_c/m_{\rm pl}^4 \sim (H_0/m_{\rm pl})^2$. Its smallness reflects the large hierarchy between the present day Hubble scale and the fundamental energy scale governing gravity.
The above result is mostly independent of the details of the inflationary model and the exact time evolution of the solutions to the equations of motion between the time of inflation and now, as in Section \ref{subsec:deriv wig} it was derived directly from the form of the action today.
As long as one operates within the assumptions stated in the beginning of this Section, there is therefore no point in exploring variations on the ultra slow-roll scenario in search of cases where the ``quantum'' decaying mode of the primordial perturbations is not hopelessly suppressed.

While the argument presented in this Section strongly constrains the possibility of having a non-negligible decaying mode signal, it is also a useful starting point for any further search of (more complex) scenarios that {\it do} predict an observable quantum remnant.
In particular, to break the assumptions that went into the above argument for the $\sim 110$ orders of magnitude suppression,
one would need to include the effects of self-interactions, mixing with additional degrees of freedom (such as additional fields during inflation or, in general, isocurvature perturbations), interactions with other environments (e.g.~the microphysical degrees of freedom describing the radiation fluid after reheating), or non-Gaussian initial states. These are all interesting directions for further study (see also existing works, e.g.~\cite{Lesgourgues:1996jc, proprigo07}), although it is not at all unlikely that the end result will still be that the decaying mode is extremely suppressed.

\subsection{Decoherence and analogy with classicality of the macroscopic world}
\label{subsec:dec}

We have shown in Section~\ref{subsec:deriv wig} that the classicality of the primordial perturbations (in the sense of being described by a very squeezed state) is related to the existence of very classical states for macroscopic objects, e.g.~for the position of an object with macroscopic mass. For any Gaussian state (i.e.~including coherent and squeezed states), the area defined by the quantum spread in phase space is determined by an effective mass parameter (by minimally realizing Heisenberg's uncertainty principle in some basis).
For a macroscopic object with mass $M$, the very large value of $M/\hbar$ in the relevant units for observations thus makes it possible to have coherent states that have unobservably small quantum spread in {\it both} directions in phase space, thus making them effectively classical. The primordial perturbations also have an extremely large effective mass, $M_\mathcal{R}/\hbar$, in the relevant units. In this case,
one direction in phase space {\it is} observably large due to squeezing. This then means that the other direction, corresponding to the ``quantum'' decaying mode, is extremely suppressed. Here the result is thus also a classical state, but in the specific sense of being extremely squeezed, which is different than the more truly classical coherent state discussed for the position of the macroscopic object.

While such a squeezed state is classical in the sense of expectation values of $n-$point functions of $\mathcal{R}$ and $\Pi$, it is very ``quantum'' in the sense that it describes a coherent quantum superposition of a macroscopically large range of values of $\mathcal{R}$. Decoherence, due to entanglement with the environment (e.g.~curvature perturbations at shorter wavelengths \cite{nelson16}), destroys the quantum coherence of this superposition and converts the pure state into a mixed state. Qualitatively, one expects decoherence of the primordial squeezed state to produce a classical (i.e.~incoherent) mixture of more or less coherent states\footnote{We note that decoherence widens the Wigner function and increases its enclosed area. Classicalization by decoherence in that sense thus has the opposite effect of the classicalization by squeezing on which we have focused in this work.}.
The phase-space area of these individual coherent states is again given by $\hbar/M_\mathcal{R}$ and thus extremely small in the relevant units for observation. Thus, after decoherence, the analogy with the classical nature of macroscopic objects is complete.

A macroscopic mass could theoretically be prepared in a Schr\"odinger's cat-like, coherent superposition of, say, two coherent states at macroscopically different positions, where each coherent state has negligible quantum spread in both directions in phase space, but it would then immediately decohere into a classical mixture of those two coherent states, destroying any possibility of seeing quantum interference between the two branches of the wave function. Similarly, the primordial curvature perturbations are originally produced in a coherent superposition of a macroscopically large range of mode amplitudes, which due to decoherence is turned into a classical mixture of coherent states. Due to the tiny value of $\hbar/M_\mathcal{R}$, these individual coherent states are extremely likely to have negligible quantum uncertainty in {\it both} directions in phase space, so that the resulting state would be truly indistinguishable from a classical distribution. When decoherence is included in the picture, the late-Universe cosmic perturbations are thus classical, macroscopic objects.
It remains to be seen if ``exotic'' scenarios exist where the above narrative breaks down and a non-negligible quantum signature remains.


\section{Summary}

In this paper, we have revisited the quantum-to-classical transition of the primordial curvature perturbations generated by inflation.
We have focused on the {\it squeezing} of the quantum state of these perturbations, a phenomenon that plays a key role in the generation of observably large curvature fluctuations and in their transition to classical behavior.
We have presented a mostly self-contained review of squeezing of the primordial perturbations, phrased explicitly in terms of the evolution of the two independent solutions to the equations of motion and the manifestation of this evolution in the Wigner function.
In this language, strong squeezing, and therefore the realization of classical behavior of expectation values, is a consequence of the decaying mode solution becoming negligible compared to the growing mode solution.
The decaying mode describes the minimal non-commuting component of the perturbations so that, in the limit where it becomes negligible, the quantum nature of the perturbations becomes hidden.

The motivating question for this work was whether there are scenarios where a remaining, explicit quantum signature survives into the late Universe where observations take place. More concretely, we asked whether the ``quantum'' decaying mode can be observable in certain cases.
We addressed this question by explicitly evolving the mode functions through the inflationary stage, the transition from inflation to the post-inflationary Universe, and finally through the post-inflationary stage until the time the perturbations have entered the horizon again in the late Universe.

We first reviewed the standard scenario of single-field, slow-roll inflation followed by a phase of radiation domination and we recovered the well known result that the decaying mode is hopelessly suppressed around the time a mode re-enters the horizon in the late Universe and is observed.
In our simple model, the numerical suppression is about $115$ orders of magnitude, leaving no hope whatsoever for observing the decaying mode in this scenario.
We then for the first time considered in detail the quantum-to-classical transition of the primordial perturbations in ultra-slow-roll inflation (again including the evolution of the state of the perturbations into the post-inflationary Universe).
The motivation for studying the ultra-slow-roll scenario is that during USR inflation, the super-horizon behavior of the growing and decaying mode solutions is radically different than in the standard slow-roll case.
After explicit calculation, we found that, while the mode evolution is indeed very different than in the SR scenario, when the perturbations are observed in the post-inflationary Universe, the suppression of the decaying mode is again given by the same $\sim 115$ orders of magnitude as in the standard slow-roll scenario.

We have finally shown that obtaining the exact same suppression factor in both scenarios is no coincidence, but is a general result related to a version of Heisenberg's uncertainty principle and the conservation of the Wronskian of the mode functions.
Assuming a Gaussian initial state, and linear evolution of the perturbations in a single component, we have shown that a decaying mode suppression of $\sim 115$ orders of magnitude is a simple consequence of the observed amplitude of the primordial power spectrum and the large hierarchy between the scale of the Universe today ($\sim H_0$) and the Planck scale.
Our results thus confirm the standard wisdom that the quantum-to-classical transition is extremely thorough and that it will be hard if not impossible to detect any remaining, explicit quantum signatures.
On the other hand, our general argument for the strong suppression of the decaying mode provides useful guidance for any further search for a primordial quantum signal.
If we want to have any chance of finding a scenario where the decaying mode is not negligible, we need to break the assumptions going into the above argument.
This motivates further study of non-Gaussian intitial states, non-linear evolution of the perturbation, and the inclusion of additional fields/fluids in the evolution.

Finally, we have briefly discussed the importance of decoherence for the quantum-to-classical transition in Section \ref{subsec:dec}.
However, the actual calculations in this paper ignore decoherence and
apply to the idealized scenario where the individual modes $\mathcal{R}_\k$ of the primordial curvature perturbations are in a pure quantum state even in the late Universe
The philosophy has been to first look for interesting scenarios in this simple, more easily computable setting.
If a model with large quantum signature had been found, it would have then made sense to make the analysis more realistic by including decoherence.

\acknowledgements

\copyright 2019. All rights reserved.  We would like to thank Chen He Heinrich, Tomislav Prokopec, J\'er\^ome Martin, Enrico Pajer, Daniel Baumann, Daniel Green, and J\'er\^ome Gleyzes for helpful discussions and gratefully acknowledge support by the Heising-Simons Foundation. Part of the research described in this paper was carried out at the Jet Propulsion Laboratory, California Institute of Technology, under a contract with the National Aeronautics and Space Administration. 



\addcontentsline{toc}{section}{References}
\bibliographystyle{utphys}
\bibliography{refs}
\end{document}